\documentclass{aa}

\usepackage{amssymb}
\usepackage[usenames,dvipsnames]{xcolor}
\usepackage{graphicx}
\usepackage{nicefrac}
\usepackage[rightcaption]{sidecap}
\usepackage{natbib}
\usepackage{txfonts}
\usepackage[scaled=.86]{helvet}
\usepackage[normalem]{ulem}
\usepackage{booktabs}
\usepackage{url}
\usepackage{hyperref}
\hypersetup{
    final=true,
    pageanchor=true,
    colorlinks=true,
    breaklinks=true,
    linkcolor=blue,
    citecolor=blue,
    urlcolor=blue,   
    pdfpagemode=UseNone,
    pdftitle={High-resolution spectroscopy of a surge in an emerging flux 
region},
    pdfauthor={M. Verma et al.},
    pdfsubject={Solar Physics},
    pdfkeywords={Sun: activity -- photosphere -- chromosphere -- Line: profiles 
    -- Methods: observational}}
    
\usepackage{pifont}
\usepackage{epstopdf}

\begin{document}


\title{High-resolution spectroscopy of a surge in an emerging flux region}

\author{%
    M.\ Verma\inst{1},  
    C.\ Denker\inst{1},
    A.\ Diercke\inst{1,2}, 
    C.\ Kuckein\inst{1},
    H.\ Balthasar\inst{1},
    E.\ Dineva\inst{1,2},\\
    I.\ Kontogiannis\inst{1},
    P.\ S.\ Pal\inst{1,3}, \and
    M.\ Sobotka\inst{4}}
    
\institute{%
    Leibniz-Institut f{\"u}r Astrophysik Potsdam (AIP),
    An der Sternwarte~16,
    14482 Potsdam,
    Germany,
    \href{mailto:mverma@aip.de}{\textsf{mverma@aip.de}}
\and
    Universit{\"a}t Potsdam,
    Institut f{\"u}r Physik und Astronomie,
    Karl-Liebknecht-Stra{\ss}e 24\,--\,25,
    14476 Potsdam,
    Germany
\and University of Delhi,
     Bhaskaracharya College of Applied Sciences,
     Sector 2, Phase 1, 
     Dwarka New Delhi-110075, 
     India
\and
    Astronomical Institute of the Czech Academy of Sciences,
    Fri\v{c}ova 298,
    25165 Ond\v{r}ejov, Czech Republic}
\authorrunning{Verma et al.}  
  
\date{Received September 23, 2019; accepted May 7, 2020}


\abstract
{}
{The regular pattern of quiet-Sun magnetic fields was disturbed by newly 
emerging magnetic flux, which led a day later to two homologous surges after 
renewed flux emergence, affecting all atmospheric layers. Hence, simultaneous 
observations in different atmospheric heights are needed to understand the 
interaction of rising flux tubes with the surrounding plasma, in particular by 
exploiting the important diagnostic capabilities provided by the strong 
chromospheric H$\alpha$ line regarding morphology and energetic processes in 
active regions.}
{A newly emerged active region NOAA~12722 was observed with the Vacuum Tower 
Telescope (VTT) at Observatorio del Teide, Tenerife, Spain on 11~September 2018. 
High-spectral resolution observations using the echelle spectrograph in the 
chromospheric H$\alpha$ $\lambda$6562.8~\AA\ line were obtained in the early 
growth phase. Noise-stripped H$\alpha$ line profiles yield maps of line-core and 
bisector velocities, which were contrasted with velocities inferred from Cloud 
Model inversions. A high-resolution imaging system recorded simultaneously 
broad- and narrow-band H$\alpha$ context images. The Solar Dynamics Observatory 
(SDO) provided additional continuum images, line-of-sight (LOS) magnetograms, 
and UV/EUV images, which link the different solar atmospheric layers.}
{The active region started as a bipolar region with continuous flux emergence 
when a new flux system emerged in the leading part during the VTT observations, 
resulting in two homologous surges. While flux cancellation at the base of the 
surges provided the energy for ejecting the cool plasma, strong proper motions 
of the leading pores changed the magnetic field topology making the region 
susceptible for surging. Despite the surge activity in the leading part, an arch 
filament system in the trailing part of the old flux remained stable. Thus, 
stable and violently expelled mass-loaded ascending magnetic structures can 
co-exist in close proximity. Investigating the height dependence of LOS 
velocities revealed the existence of neighboring strong up- and downflows. 
However, downflows occur with a time lag. The opacity of the ejected cool plasma 
decreases with distance from the base of the surge while the speed of the ejecta 
increases. The location at which the surge becomes invisible in H$\alpha$ 
corresponds to the interface where the surge brightens in He\,\textsc{ii} 
$\lambda$304~\AA. Broad-shoulders and dual-lobed H$\alpha$ profiles suggests 
accelerated/decelerated and highly structured LOS plasma flows. Significantly 
broadened H$\alpha$ profiles imply significant heating at the base of the 
surges, which is also supported by bright kernels in UV/EUV images uncovered by 
swaying motions of dark fibrils at the base of the surges.}
{The interaction of newly emerging flux with pre-existing flux concentrations of 
a young, diffuse active region provided suitable conditions for two homologous 
surges. High-resolution spectroscopy revealed broadened and dual-lobed H$\alpha$ 
profiles tracing accelerated/decelerated flows of cool plasma along the 
multi-threaded structure of the surge.}

\keywords{Sun: activity -- photosphere -- chromosphere -- Line: profiles 
    -- Methods: observational}

\maketitle


\section{Introduction}\label{SEC01}

The chromosphere is a very dynamic layer of the solar atmosphere, which displays 
various fine structures on different spatial and temporal scales 
\citep[e.g.,][]{Beckers1964, Bray1974}. A surge is a chromospheric eruptive 
feature visible in the blue and red wings of the H$\alpha$ line, where cool 
plasma is ejected. The first surge observations were reported in the early 1940s 
\citep[e.g.,][]{Newton1942, Ellison1942}. Since then, various properties of 
H$\alpha$ surges were at the focus of many studies. Apart from H$\alpha$, surges 
are visible in other chromospheric lines, for example, in He\,\textsc{ii} 
304~\AA\ \citep{Georgakilas1999}, H$\beta$ \citep{Zhang2000}, Ca\,\textsc{ii}\,H 
\citep{Tziotziou2005}, C\,\textsc{iv}, and several UV/EUV lines 
\citep[e.g.,][]{Liu2004, Kayshap2013}.

\citet{Bruzek1974} explained surges as curved or straight spikes of material 
shooting out, reaching coronal heights at velocities of up to 200~km~s$^{-1}$ 
with a lifetime of 10\,--\,20~min. The surge material appears usually in 
absorption on the solar disk. \citet{Roy1973a} suggested that surges share 
properties with prominences, where material is supported by the strong magnetic 
fields above active regions. \citet{Roy1973b} noted that surges are expected in 
the proximity of spots or pores with small-scale flares or brightenings such as 
Ellerman bombs \citep{Ellerman1917}. In addition, regions with evolving magnetic 
features, that is, most notably emerging flux regions (EFRs), are more prone to 
have surges. \citet{Roy1973a} found that the velocity curve for surges begins 
with material being accelerated to maximum velocity, and then decelerates under 
the influence of the magnetic field supporting the surge.

Magnetic reconnection was proposed by \citet{Yokoyama1995} as the origin of 
H$\alpha$ surges. In their simulations, the whip-like motion of cool 
chromospheric plasma is produced because of reconnection in an EFR. When the 
cool plasma is carried to higher atmospheric layers by the expanding flux loops, 
reconnection in these loops produces the slingshot-like motion, which they 
associated with H$\alpha$ surges. Small-scale magnetic features and their 
association with surges were the focus of the investigation by 
\citet{Brooks2007} who linked flux cancellation in an emerging flux region to 
surges. Slow reconnection related to flux cancellation was identified by 
\citet{Chae1999} as the cause of a surge, where the ejection of cool and hot 
plasma along different magnetic field lines leads to signatures in H$\alpha$ and 
EUV lines, respectively. The base of surges and kernels of UV brightenings are 
often located near magnetic neutral lines, where opposite-polarity features 
collide \citep{Yoshimura2003}.

Spectral observations of surges were presented in many studies, for example, 
\citet{Schmieder1983} selected the H$\alpha$ and C\,\textsc{iv} lines for their 
observations. H$\alpha$ showed lower acceleration, and a shock wave appeared in 
the upper part of the surge. \citet{Schmieder1994} addressed the question what 
mechanism is responsible for driving the surge and maintaining the cool plasma 
ejection. By comparing two transient events in the same active region, they 
explored the ejection mechanism, and suggested that just a pressure-driven 
mechanism is insufficient for explaining the properties of the ejected material. 
\citet{Madjarska2009} discussed trigger mechanisms of a surge using filtergrams 
taken at five positions along the H$\alpha$ line. The observed surge seemed to 
be triggered by an Ellerman bomb. The authors emphasized the need for 
high-spatial as well as high-spectral resolution to disentangle various 
chromospheric phenomena. 

\citet{NobregaSiverio2016} presented numerical modeling of a surge and performed 
a 2.5D numerical experiment describing the emergence of magnetized plasma from 
lower atmospheric layers to the corona. They considered entropy sources, which 
allowed them to trace the evolutionary pattern of the plasma elements in the 
surge. They concluded that entropy sources are essential to understand thermal 
properties of surges. Combining radiation transfer 
\citep[Bifrost,][]{Gudiksen2011} and extensive Lagrange tracing, they identified 
four different populations in the surge. One population covers 34\% of the 
surge's cross-section and reaches temperatures as high as 10$^6$~K, which 
however returns to typical surge temperatures of 10$^4$~K by radiative losses 
and thermal conduction. This suggests that heating and cooling processes are 
important aspects of the surging plasma. In \citet{NobregaSiverio2017, 
NobregaSiverio2018}, they extended the numerical experiments to Si\,\textsc{iv} 
and O\,\textsc{iv} lines to understand the flow of material and energy in the 
transition region.

The strong absorption line H$\alpha$ provides access to the thermal and dynamic 
properties of chromospheric features. For example, \citet{Verma2012a} used 
high-resolution H$\alpha$ spectra in combination with data of the Japanese 
\textit{Hinode} space mission \citep{Kosugi2007} to study the slow decay of two 
spots over five days, and \citet{Kuckein2016} studied a large quiet-Sun filament 
using similar H$\alpha$ spectroscopic data. The potential of a dedicated data 
pipeline for bulk-processing of H$\alpha$ spectra is presented in 
\citet{Dineva2020}, where noise-stripping based on Principal Component Analysis 
(PCA) prepares H$\alpha$ contrast profiles for Cloud Model 
\citep[CM,][]{Beckers1964} inversions. 

In the present work, we follow the evolution of an EFR with strong absorption 
features in proximity to sites of continuous flux emergence. In 
Sect.~\ref{SEC02}, we present the observations and the various steps to process 
the data. In Sect.~\ref{SEC03}, we introduce slit-reconstructed intensity and 
velocity maps and describe the temporal and spatial variation of H$\alpha$ line 
profiles that are associated with the two homologous surges. In addition, we 
discuss the magnetic field evolution and the response of the upper atmosphere. 
Our results are discussed in Sect.~\ref{SEC04} and placed in the context of 
relevant literature related to surges and other ejecta. The high-spectral and 
good spatial resolution of the Vacuum Tower Telescope 
\citep[VTT,][]{vonderLuehe1998} echelle spectrograph were the point of departure 
for this work and allowed us to investigate the complex evolution and dynamics 
of the strong absorption features for almost four hours -- thus adding to the 
still sparse knowledge of the spectral characteristics of surges. Future work 
and prospects for high-cadence scans with the VTT echelle spectrograph are 
described in our concluding remarks in Sect.~\ref{SEC05}.


\section{Observations and data reduction}\label{SEC02}

The spectroscopic data covering active region NOAA~12722 were acquired at the 
VTT on 11~September 2018. In total 20 scans were taken in three spectral lines 
between 08:05~UT and 11:49~UT, describing the evolution of the active region. 
During this time period of about 3.5~hours, only two gaps of 20 and 17~min 
occurred between scans 9 \& 10 and 15 \& 16, respectively. Five sets of flat 
fields frames with 200 scan steps were recorded at solar disk center while 
changing the telescope pointing following a circular track. The average dark 
frame was computed from 200 individual frames.

The spectral lines H$\alpha$ $\lambda$6562.8~\AA, Cr\,\textsc{i} 
$\lambda$5781~\AA, and H$\beta$ $\lambda$4861~\AA\ were simultaneously recorded 
with the VTT echelle spectrograph, utilizing the chromospheric grating with a 
blaze angle of 62\degr. Even though all the three spectral lines were processed, 
the present study focuses solely on H$\alpha$ spectroscopy. The slit width was 
80~$\mu$m, resulting in an exposure time of 300~ms, which matches the full-well 
capacity of the detector. A broad-band interference filter with a 
full-width-at-half-maximum (FWHM) of 8.7~\AA\ and with a peak transmission of 
62\% was used to suppress overlapping spectral orders. The pco.4000 CCD camera 
has a quantum efficiency of about 30\,--\,40\% in the spectral range of 
4000\,--\,6500~\AA. After 2$\times$2-pixel binning and discarding the borders, 
the spectra have a size of 2000 $\times$ 660 pixels. The pixel size of the CCD 
detectors is 9~$\mu$m $\times$ 9~$\mu$m. The resulting pixel size on the solar 
surface is about 0.18\arcsec\ in the slit direction. The scan step size is 
0.16\arcsec, and 630 scan steps are recorded, scanning a range of 
$\pm$50\arcsec\ around the lock point of the Kiepenheuer Adaptive Optics System 
\citep[KAOS,][]{vonderLuehe2003}. This scan sequence takes about 9~min and 
results in slit-reconstructed maps with a size of 100\arcsec $\times$ 
120\arcsec\ for the physical parameters derived from the spectral profiles. The 
dispersion is about 4.2~m\AA~pixel$^{-1}$, and the spectral range covered by the 
detector is about 8.4~\AA.

The spectral data went through the usual preprocessing steps, which includes 
dark subtraction, flat-fielding, removal of the spectrograph profiles, etc. 
Noise is removed from the H$\alpha$ spectra using PCA, and CM inversions are 
computed for all the 20 scans. The details of the pre-processing steps and CM 
inversions are presented in \citet{Dineva2020}, who used iterative PCA for 
dimensionality reduction, resorting to only ten eigenfunctions, which represent 
the H$\alpha$ spectral line. The resulting spectra are resampled to only 601 
wavelength points, which cover a wavelength region of $\pm$3~\AA\ around the 
H$\alpha$ line core. The coarser dispersion of 10~m\AA~pixel$^{-1}$ 
significantly reduced the computation time for CM inversions and PCA 
decomposition. The final results are noise-free spectra and maps of physical 
parameters computed using these profiles. An example comparing observed 
intensity and contrast profiles with the noise-free spectra is compiled in 
Fig.~7 of \citet{Dineva2020}. The CM inversions delivered four physical 
parameters: optical thickness $\tau$, Doppler width $\Delta\lambda_D$, 
line-of-sight (LOS) velocity of the cloud material $v_D$, and source function 
$S$.

\begin{figure}[t]
\centering
\includegraphics[width=\columnwidth]{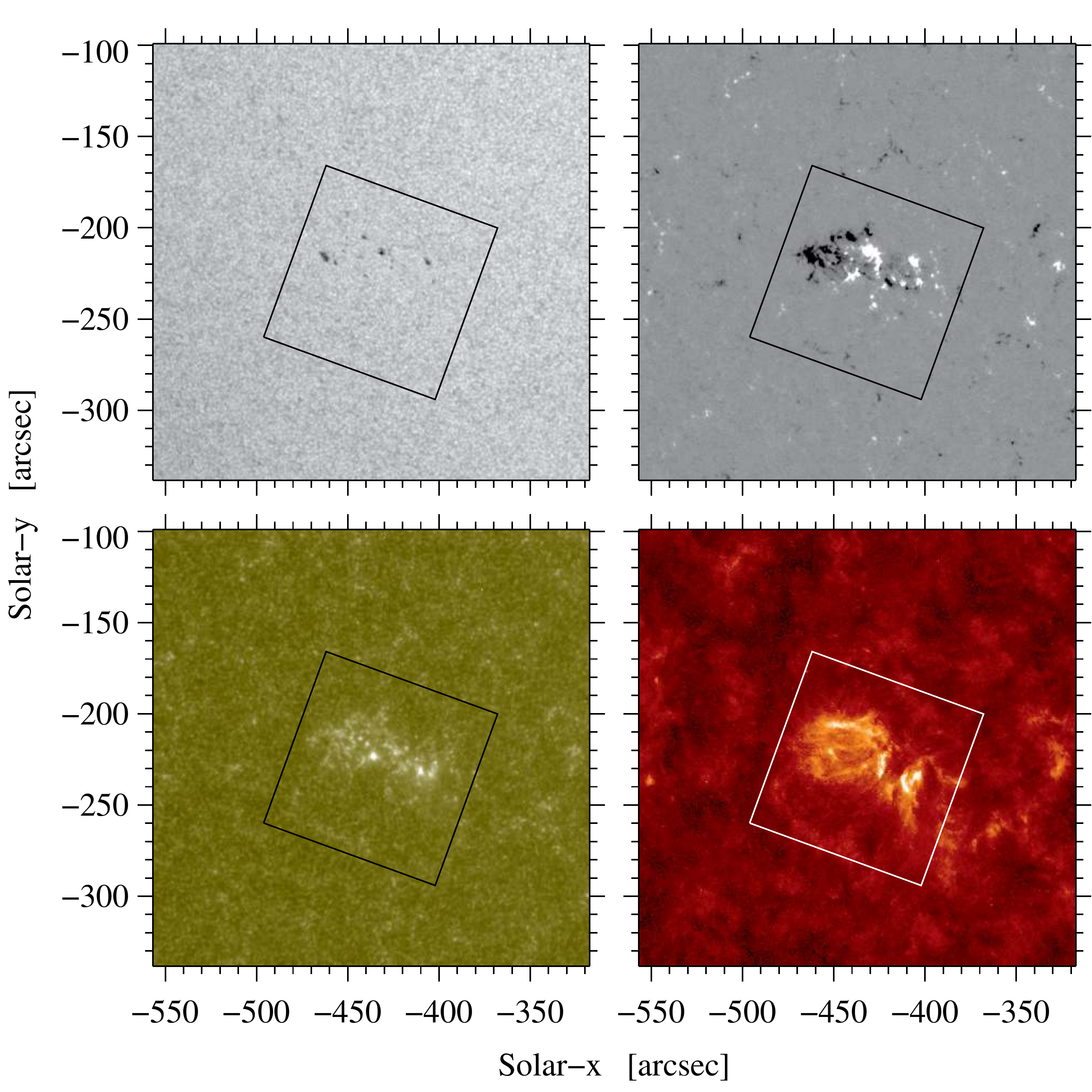}
\caption{Overview of active region NOAA~12722: HMI continuum image
    (\textit{top-left}), HMI magnetogram (\textit{top-right}), AIA UV
    $\lambda$1600~\AA\ image (\textit{bottom-left}), and AIA EUV He\,\textsc{ii}
    $\lambda$304~\AA\ image (\textit{bottom-right}) observed at 08:05~UT on 
    11~September~2018. The magnetogram was clipped between $\pm$250~G. The 
    black and white boxes represent the central $100\arcsec \times 100\arcsec$ 
    of the FOV scanned by the VTT echelle spectrograph (see online movie with a 
one-minute cadence for a detailed view).} 
\label{FIG01}
\end{figure}

As an alternative to observations with the echelle spectrograph, a system of two 
synchronized LaVision Imager M-lite 2M CMOS cameras were used to capture narrow- 
and broad-band H$\alpha$ filtergrams. The narrow-band filtergrams were obtained 
with a Lyot filter ($\lambda$6562.8~\AA\ and $\Delta\lambda = 0.60$~\AA) 
manufactured by Bernhard Halle Nachf., Berlin-Steglitz \citep{Kuenzel1955}, and 
the broad-band filtergrams were captured with an interference filter 
($\lambda$6567~\AA\ and $\Delta\lambda = 7.5$~\AA). The details about setup and 
data processing are given in \citet{Denker2020}. Note that these images were not 
simultaneously taken with spectral data but earlier at 07:50~UT. Here, the 
restored narrow- and broad-band filtergrams serve as high-resolution context 
images. Images were restored using Multi-Object Multi-Frame Blind Deconvolution 
\citep[MOMFBD,][]{vanNoort2005}. 

The Solar Dynamics Observatory \citep[SDO,][]{Pesnell2012} provided full-disk 
context images and magnetograms for discussing the temporal evolution and 
morphology of the active region. The continuum images and LOS magnetograms were 
obtained with the Helioseismic and Magnetic Imager \citep[HMI,][]{Scherrer2012}, 
whereas the Atmospheric Imaging Assembly \citep[AIA,][]{Lemen2012} provided UV 
$\lambda$1600~\AA\ and EUV He\,\textsc{ii} $\lambda$304~\AA\ images. These two 
wavelengths cover different layers in solar atmosphere with about 430~km 
\citep{Fossum2005} and about 4000~km \citep{Alissandrakis2019} as formation 
heights, respectively. Data processing follows the steps outlined in 
\citet{Beauregard2012} and \citet{Verma2018a}.

\begin{figure}[t]
\centering
\includegraphics[width=\columnwidth]{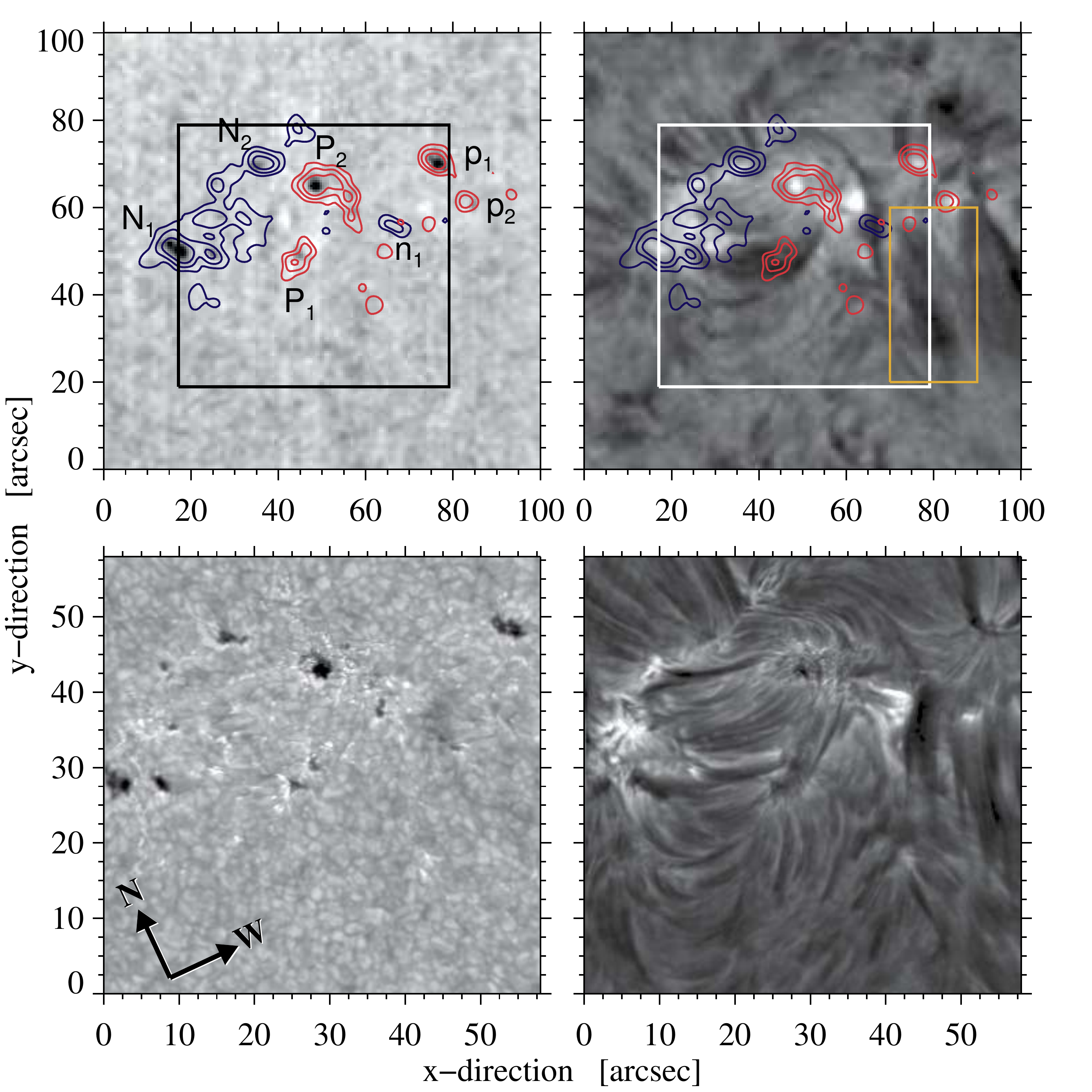}
\caption{Slit-reconstructed H$\alpha$ continuum (\textit{top-left}) and 
line-core 
    (\textit{top-right}) intensity images of active region NOAA~12722 observed 
    at 08:05~UT on 11~September 2018. The blue and red contours refer to a
    co-temporal magnetogram at $\pm$100, $\pm$250, and $\pm$500~G. The black
    and white squares mark the location of restored H$\alpha$ broad-band
    (\textit{bottom-left}) and narrow-band line-core (\textit{bottom-right})
    filtergrams taken at 07:50~UT. The yellow rectangle (\textit{top-right}) 
    marks the location of continuous surge activity.} 
\label{FIG02}
\end{figure}


\section{Results}\label{SEC03}

Active region NOAA~12722 appeared on the solar surface at around 19:00~UT on 
10~September 2018. By the time of the VTT observations on 11~September 2018, the 
region (see Fig.~\ref{FIG01}) was located at disk-center coordinates 
(440\arcsec~E, 210\arcsec~S), which correspond to a cosine of the heliocentric 
angle of $\mu \approx 0.86$. The full field-of-view (FOV) covered by the scans 
with the VTT echelle spectrograph was cropped to $100\arcsec \times 100\arcsec$ 
as indicated by black and white square boxes in Fig.~\ref{FIG01}. A 
corresponding movie with the detailed view of the central $100\arcsec \times 
100\arcsec$ is presented as online material. Since AIA and HMI data are recorded 
at different cadences from 12~s to 45~s, the time-series were resampled to a 
one-minute cadence covering the period 07:30\,--\,12:00~UT. For this purpose, 
the time-series were corrected for solar differential rotation before using 
linear interpolation to an equidistant grid in time on a pixel-by-pixel basis. 
This restricted FOV as presented in the top panels of Fig.~\ref{FIG02} will be 
used in the following for the investigation of the surges.

In the next sections, we describe first the magnetic field evolution and 
response of the upper atmosphere based on the time-series of SDO data and then 
study various line properties extracted from the noise-free H$\alpha$ spectra. 
The morphology of absorption features was derived from slit-reconstructed red- 
and blue-wing as well as line-core maps. Time-series of matching LOS velocity 
maps reveal the dynamics of the surge, while selected H$\alpha$ line profiles 
allow us to describe their complex spatio-temporal properties.


\subsection{Magnetic field evolution}\label{SEC31}

The active region emerged with the trailing negative polarity being more 
prominent with two tiny pores and a very dispersed leading positive polarity, 
which deviates from the typical emergence pattern of active regions. Tiny pores 
of positive polarity appeared at around 05:00~UT on 11~September 2018. In 
addition, a new flux system with mixed polarity started to emerge in the leading 
part of the already existing region at around 06:00~UT. When the VTT 
observations started at 08:05~UT, the region consisted of multiple pores with 
continuous flux emergence in the leading region as shown in Fig.~\ref{FIG01}. In 
Fig.~\ref{FIG02}, the slit-reconstructed continuum and line-core maps from the 
first scan are displayed with contours of a magnetogram taken at the same time. 
The pores belong to the already existing system with flux emergence in the 
trailing part and are marked as \textsf{P}$_\mathsf{1}$, 
\textsf{P}$_\mathsf{2}$, \textsf{N}$_\mathsf{1}$, and \textsf{N}$_\mathsf{2}$ 
for positive and negative polarities, respectively. The pores associated with 
the new flux system are labeled by \textsf{p}$_\mathsf{1}$, 
\textsf{p}$_\mathsf{2}$, and \textsf{n}$_\mathsf{1}$, referring again to 
positive and negative polarities. Several other features with mixed polarities 
are not discretely tagged.

\begin{figure}[t]
\centering
\includegraphics[width=\columnwidth]{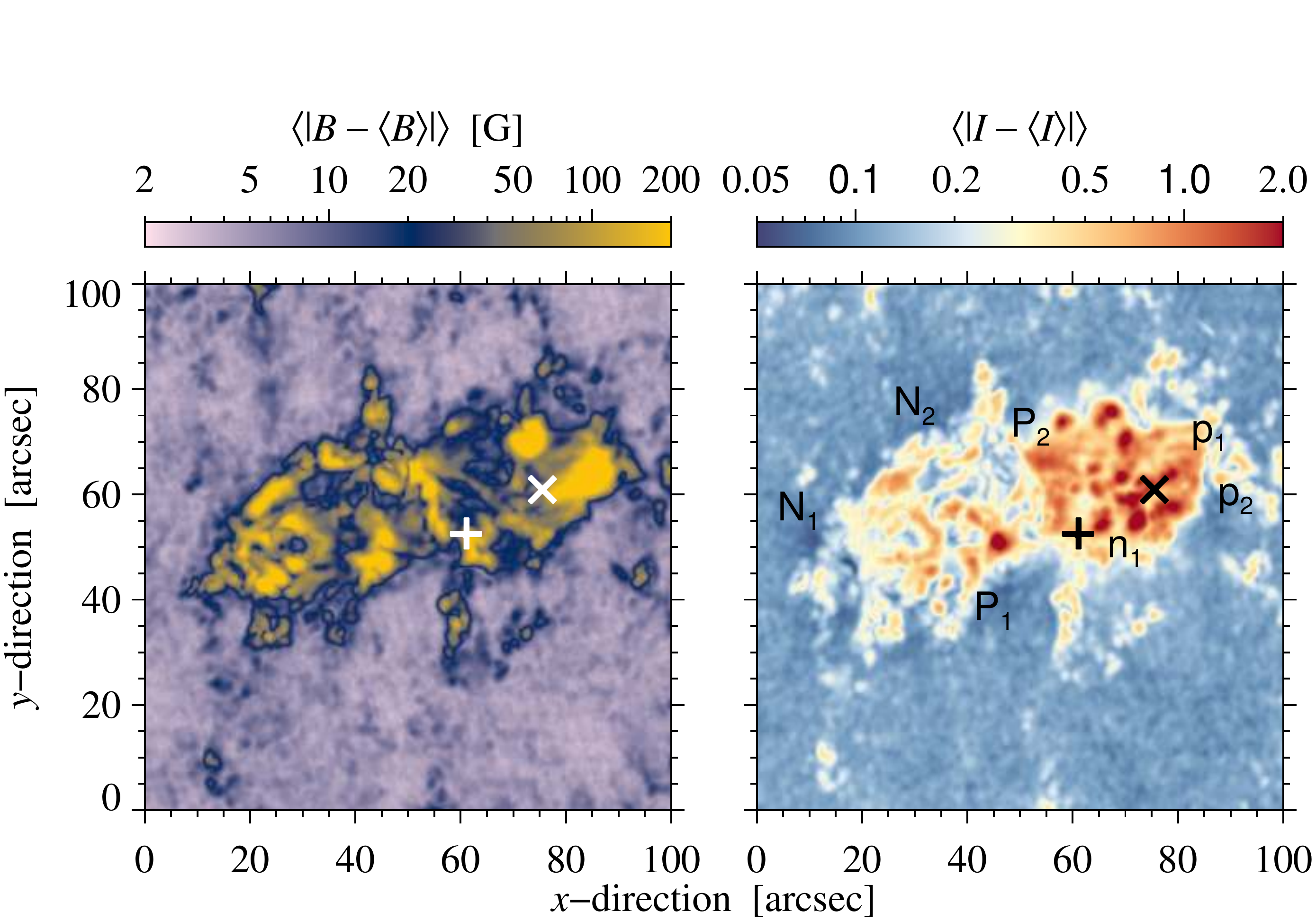}
\caption{Background-subtracted activity maps of the LOS magnetic field $B$ 
(\textit{left}) and the UV $\lambda$1600~nm intensity $I$ (\textit{right}). The 
maps were computed for the time period 07:30\,--\,12:00~UT and show the 
locations with strong variations of the magnetic field and UV intensity. The 
black `\textsf{+}' and `$\times$' mark the location of persistent blue- and 
redshifted regions as discussed in Sect.~\ref{SEC35} and depicted in 
Fig.~\ref{FIG05}. The labels in the right panel are the same as in 
Fig.~\ref{FIG02}.} 
\label{FIG03}
\end{figure}

The restored high-resolution images from the CMOS cameras show 
diffraction-limited details of the photosphere and chromosphere (bottom panels 
of Fig.~\ref{FIG02}). Elongated granules between pores \textsf{P}$_\mathsf{1}$ 
and \textsf{N}$_\mathsf{2}$ indicate on-going flux emergence 
\citep[e.g.,][]{Verma2016a}. In addition, the H$\alpha$ broad-band image and the 
time-series of HMI continuum images show both indications of thread-like dark 
intergranular lanes connecting opposite magnetic polarities 
\citep[e.g.,][]{Strous1996}, which are a signature of emerging $\Omega$-loops 
capable of lifting cool plasma into higher atmospheric layers. As a result, a 
typical arch filament system \citep[AFS, e.g.,][]{Sergio2018} is evident in the 
H$\alpha$ narrow-band filtergram between these pores. The pre-existing flux in 
the trailing bipolar region evolves in general gradually, leading to a 
simplification and stretching of the active region in the trailing part over 
time.

In contrast, the newly emerging flux system exhibits strong upwelling flux to 
the north between pores \textsf{P}$_\mathsf{2}$ and \textsf{p}$_\mathsf{1}$ in 
form of small-scale negative-polarity features resembling type~III moving 
magnetic features \citep[MMFs,][]{Kubo2007b} but in absence of a moat region 
around the pores. The observed enhanced H$\alpha$ line-wing and line-core 
emission, that is, bright H$\alpha$ grains, may be signatures of reconnection 
events of undulating, serpentine-like field lines, which rise up in the 
atmosphere as mass-loaded \textsf{U}-loops \citep{Pariat2004}. These bright 
H$\alpha$ grains coincided with UV brigthenings, similar to UV bursts which were
associated by \citet{Ortiz2020} with Ellerman bombs and emerging flux. 
Strengthening of pores \textsf{p}$_\mathsf{1}$ and \textsf{p}$_\mathsf{2}$ 
balances the flux emergence. The significant growth and strong proper motion 
towards the west of pore \textsf{p}$_\mathsf{2}$ impacts the magnetic field 
topology of the active region and is besides flux cancellation an ingredient for 
surging. The base of the surge is associated with the negative-polarity feature 
\textsf{n}$_\mathsf{1}$ in the western part of the region but not exactly at the 
periphery \citep[cf.,][]{Mulay2016}. Here, continuous flux cancellation occurs 
during the 3\nicefrac{1}{2}-hour period covered by the SDO HMI/AIA movie. The 
cool plasma was ejected towards the south but the exact orientation of the 
surges was difficult to determine because the region is not located at the solar 
disk center, and it appears as if the material is not ejected radially. Thus, 
the LOS velocities computed in later sections may underestimate the speed of the 
ejected plasma. The broad surge-like structure is marked by the yellow rectangle 
in the upper panel of Fig.~\ref{FIG02}.
\newline


\subsection{Response of the upper atmosphere}\label{SEC32}

In this section, we discuss the response of the upper atmosphere of the surge 
activity using the online movie accompanying Fig.~\ref{FIG01}. The first 
indication of surge activity is already visible at the location of the 
cancelling magnetic feature \textsf{n}$_\mathsf{1}$ in AIA UV and EUV images at 
around 08:37~UT. The UV images show a confined brightening of two small kernels, 
and the EUV intensity lags by about one minute and reaches its peak brightening 
at around 08:40~UT. The EUV brightenings are more structured and cover a larger 
area. These UV/EUV brightenings are indicative of localized heating in the lower 
atmosphere. In the EUV images, a surge appears around this time, first as a dark 
structure and then brightens, including a thin collimated bright structure, as 
the surge progresses. The surge continues beyond the southern edge of the FOV 
(at around 08:44~UT) in the online movie belonging to Fig.~\ref{FIG01}. In the 
UV images, this surge appears only as faint hazy bright structure in proximity 
to the negative-polarity feature \textsf{n}$_\mathsf{1}$.

Starting around 09:05~UT, UV/EUV brightenings appear in the neighborhood of the 
negative-polarity feature \textsf{n}$_\mathsf{1}$. They are not always visible 
in the EUV images, where they are often obscured by opaque dark filaments higher 
up. However, the swaying motion of the filaments uncover these brightenings from 
time to time, indicating that the heating observed in the UV/EUV images occurs 
below the filamentary structure. The second major surge starts at around 
09:17~UT with a homologous appearance and propagation pattern as its precursor, 
however, affecting a much larger area with respect to brightenings and ejected 
material. Even the AFS in the trailing part of the active region is affected, 
that is, the loop system connecting opposite polarities is activated and 
significantly brightens. Again, opaque material jutting out to the south from 
the base of the surge becomes brighter and less opaque while being ejected, 
which is evident in the EUV images at around 09:36~UT. After this time some 
ejected material returns, and both bright loops and dark filaments exhibit 
complex and dynamic interactions until the end of the SDO HMI/AIA movie at 
12:00~UT.

\begin{figure*}[t]
\centering
\includegraphics[width=\textwidth]{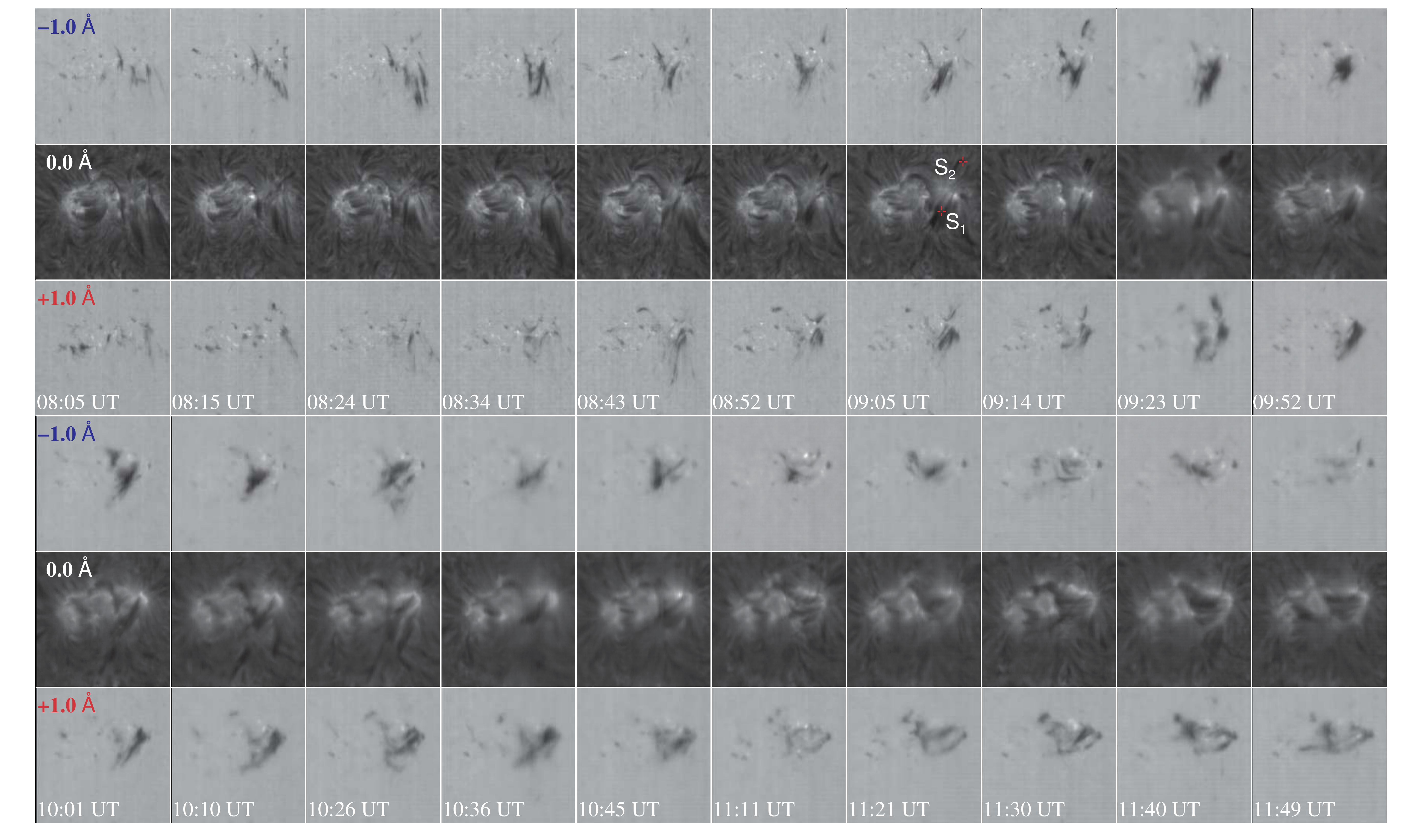}
\caption{Slit-reconstructed H$\alpha$ line-core intensity maps 
    (\textit{rows~2 and~5}) for all 20 scans along with the corresponding 
    maps of the blue (\textit{rows~1 and~4}) and red (\textit{rows~3 and~6}) 
    line-wing intensity at $\pm$1.0~\AA\ observed on 11~September 2018. 
    The image contrast of the line-core, blue and red line-wing intensity maps 
ranges between 0.15\,--\,0.80, 0.25\,--\,1.4, and 0.25\,--\,1.4 
    $I / I_0$, respectively. All maps depict the ROI of 100\arcsec $\times$ 
100\arcsec\ shown in the top panels of Fig.~\ref{FIG02}. Two locations along the 
surge are 
    indicated in panel seven of row~2 as \textsf{S}$_\mathsf{1}$ and 
\textsf{S}$_\mathsf{2}$, which are discussed in Sect.~\ref{SEC36} and depicted 
in Fig.~\ref{FIG10}. The accompanying online movie furnishes a detailed view of 
ongoing dynamics in the active region.} 
\label{FIG04}
\end{figure*}

Reconnection at the cancellation site provides the energy for propelling the 
surging cool plasma. Proper motions and shear are important ingredients to store 
energy in non-potential magnetic field configurations. In addition, they can 
open or stretch the overarching magnetic field loops allowing the plasma to 
escape. The role of shear flows and other types of proper motions has recently 
attracted renewed interest in the context of surges \citep[e.g.,][]{Yang2019}. 
The background-subtracted activity maps \citep[BaSAMs,][]{Denker2019b} in 
Fig.~\ref{FIG03} highlight locations in the active region with strong variations 
of the magnetic field and UV intensity. Continued flux emergence is clearly 
evident in the trailing bipolar part of the active region but also cancellation 
near the negative-polarity feature \textsf{n}$_\mathsf{1}$. Growth of both pores 
\textsf{p}$_\mathsf{1}$ and \textsf{p}$_\mathsf{2}$ leaves a strong signature in 
the magnetic BaSAM, which is enhanced by the strong proper motions of pore 
\textsf{p}$_\mathsf{2}$. The UV intensity variation caused by intermittent 
bightenings shows a strong disparity between the pre-existing flux system in the 
trailing part and the very dynamically evolving leading part of the active 
region. Several kernels of enhanced UV variation are visible near the 
negative-polarity feature \textsf{n}$_\mathsf{1}$ and in the `wake' of the fast 
moving pore \textsf{p}$_\mathsf{2}$.


\subsection{Temporal evolution in H$\alpha$ line-wing and line-core maps}

Blue and red line-wing along with line-core maps were created (Fig.~\ref{FIG04}) 
to investigate the evolution of the active region in different atmospheric 
heights. The H$\alpha$ line-wing maps were obtained at $\pm$1.0~\AA\ from the 
central line-core position. All three parameters are also compiled in an online 
movie, where the ongoing dynamics in the region can be visualized.

In the first two scans, contrast features appear at small scales. The 
blueshifted absorption features are more concentrated in the newly emerging 
region. The location of these features coincides with the mixed-polarity region. 
By the third and fourth scan at 08:24 and 08:34~UT, respectively, these 
absorption features become more elongated in the blue wing, which leads up to 
the first smaller surge. However, small-scale counter parts in the red wing 
remain faint. Starting with the fifth scan at 08:43~UT, extended absorption 
features appear in the red wing between pore \textsf{p}$_\mathsf{1}$ and the 
negative-polarity feature \textsf{n}$_\mathsf{1}$. The second stronger surge at 
09:05~UT led to darker absorption features at the same location, which only 
faded after 09:52~UT. By this time the features in red wing are elongated and 
located just next to the blue-wing features. The absorption features in the blue 
wing slowly disappeared and by the end of a scan at 11:49~UT, only very faint 
signatures were left, whereas the red-wing features persisted and traced out a 
bright oval-like structure in the line-core intensity maps. In general, 
absorption features in the red wing appear later than in the blue wing. This 
agrees with the findings of \citet{Yang2019} who observed that blueshifts occur 
first but that the phase of redshifts persists longer. Note that the scans take 
about 9~min, which has to be taken into account when comparing 
slit-reconstructed VTT data to more instantaneous SDO imaging/magnetogram data. 
However, certain locations in the slit-reconstructed data can be dated more 
precisely depending on the number of covered slit positions, that is, 70 slit 
positions or about 11\arcsec\ are covered in one minute. In this time interval 
about three matching He\,\textsc{ii} $\lambda$304~\AA\ images are available. 
Considering these limitations, it can be concluded that the surges appear 
simultaneously in H$\alpha$ and He\,\textsc{ii} $\lambda$304~\AA.

\begin{figure*}[t]
\centering
\includegraphics[width=\textwidth]{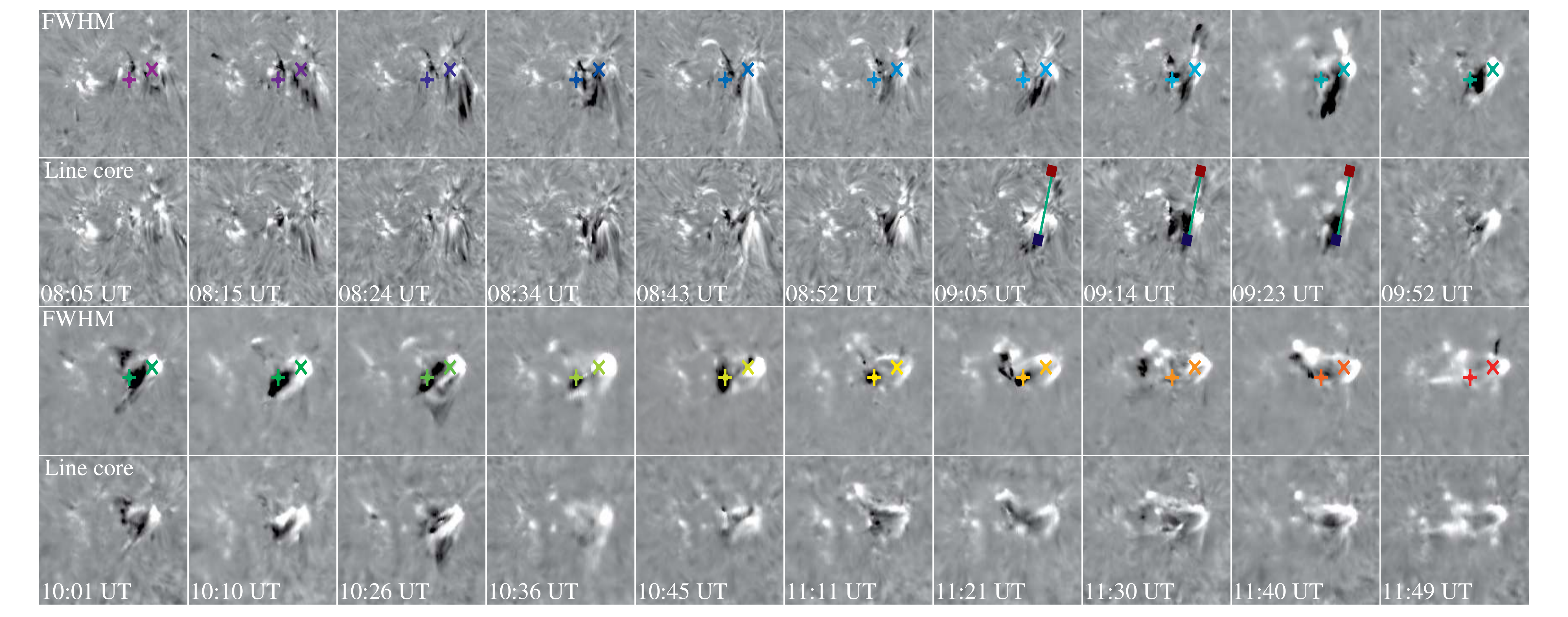}
\caption{Slit-reconstructed FWHM (\textit{rows 1 and 3}) and line-core 
    (\textit{rows 2 and 4}) velocity maps. The maps are 
    scaled between $\pm$10~km s$^{-1}$. Black and white colors correspond to 
upflows and downflows, respectively. The rainbow colored `\textsf{+}' and 
`$\times$' are persistent blue- and redshifted regions recognized by using the 
average of 20 velocity maps with LOS velocities of $\pm$3~km s$^{-1}$. The 
corresponding H$\alpha$ profiles are shown in Fig.~\ref{FIG06}. The intensity 
and contrast profiles along the green lines in panels 7\,--\,9 of are displayed 
in Fig.~\ref{FIG08}. The accompanying online movie provides a visualization of 
temporal changes in these two parameters.} 
\label{FIG05}
\end{figure*}

The two homologous surges, appearing in the line-core maps at 08:34~UT and 
09:05~UT, were observed in the newly emerging region with mixed polarities. The 
surging activity lasted longer (more than 90~min) compared to typical H$\alpha$ 
surges, which are rather short-lived. The second stronger surge extended for 
about 28\arcsec\ and was about 12\arcsec\ wide. It is not a uniform absorption 
feature but exhibits substructures with a width of 4\,--\,8\arcsec. The 
multithreaded morphology of surges may be explained by simulations of 
chromospheric ejecta along twisted magnetic field lines by \citet{Iijima2017}. 
In addition, the later stages of the active region evolution is dominated by 
downward drainage of plasma. Furthermore, small-scale brightenings are 
continuously present in line-core, blue-wing, and red-wing maps. They are 
related to bright H$\alpha$ grains caused by strong-field magnetic 
concentrations \citep[MCs,][]{Rutten2013}, which are associated with the 
upwelling magnetic flux between pores \textsf{P}$_\mathsf{2}$ and 
\textsf{p}$_\mathsf{1}$ (see Sect.~\ref{SEC31}). Only resolved spectra can 
reveal the differences between bright H$\alpha$ grains and Ellerman bombs, where 
in the latter case, the line-core intensity is not enhanced. Thus, 
misclassification may be an issue in studies resorting to H$\alpha$ filtergrams. 
Considering the dynamic flux emergence, Ellerman bombs as well as bright 
H$\alpha$ grains are likely associated with photospheric reconnection leading to 
surges. Surges and Ellerman bombs, and also bright H$\alpha$ grains, can occur 
together but not exclusively \citep[e.g.,][]{Watanabe2011}. However, the nature 
of continuously evolving small-scale mixed polarities and their ambiguous 
relationship to brightenings at various atmospheric heights hamper a clear 
causation among various features, at least in crowded and complex magnetic 
environments.


\subsection{Line-of-sight velocities}

The detailed inspection of absorption features in the red and blue wings is 
based on LOS velocity maps for the line core for all 20 scans. The line-core 
velocities are estimated by first finding the minimum of the H$\alpha$ line-core 
intensity and its position, followed by fitting a range of $\pm$0.3~\AA\ around 
this location with a parabola. Thus, the line-core velocity can be determined 
across almost the entire spectral range covered by the detector. In addition, 
similar maps of the bisector velocity corresponding to the FWHM of the H$\alpha$ 
line profile provide information of plasma motions lower in the atmosphere. The 
FWHM and the corresponding velocity were determined individually for each 
profile between minimum and maximum intensity in the observed spectral range of 
$\pm$3~\AA. The maximum intensity is almost the same in all cases, that is, 
$I_\mathrm{max} \approx 0.85\,I_0$, as a result of normalizing the spectra to 
the quiet-Sun continuum intensity.

According to \citet{Vernazza1973}, the H$\alpha$ line is formed at about 
300\,--\,1600~km in the atmosphere. As it is widely accepted, the line core 
originates higher in the chromosphere than the line wings \citep{Cauzzi2009}. 
Accurate calculations of the H$\alpha$ formation height are provided by 
\citet{Leenaarts2006a}. In their detailed work regarding the formation of the 
H$\alpha$ line, \citet{Leenaarts2012} found that the H$\alpha$ line-core 
intensity is correlated with the average formation height, that is, the 
intensity will decrease with increasing formation height. Nevertheless, 
pin-pointing the exact formation height of H$\alpha$ remains elusive. Maps of 
the FWHM and line-core velocity are compared in Fig.~\ref{FIG05}, where 
velocities are scaled between $\pm$10~km~s$^{-1}$ and red- and blueshifts (down- 
and upflows) are displayed in white and black, respectively. Noise-free spectral 
profiles after PCA decomposition and without the contamination of line blends 
helped in creating these maps. The changes in these two parameters can also be 
visualized in the accompanying online movie.

At a first glance, the velocity pattern at both heights appears to be similar. 
However, closer inspection reveals more details in the line-core maps. The most 
conspicuous changes of the LOS velocities are located in the leading part of the 
active region, where magnetic flux continuously emerges, as it is evident in all 
20 scans. Flux emergence was accompanied with small patches of downflows between 
the leading and trailing pores \textsf{P}$_\mathsf{2}$, \textsf{p}$_\mathsf{1}$, 
\textsf{p}$_\mathsf{2}$, and \textsf{n}$_\mathsf{1}$. By the second scan, 
additional small upflow patches materialized in proximity to these downflow 
patches, which are most prominent around the leading pores 
\textsf{p}$_\mathsf{1}$ and \textsf{p}$_\mathsf{2}$. The velocity pattern 
associated with the region of continuous flux emergence between the pore 
\textsf{p}$_\mathsf{1}$ and the negative-polarity feature 
\textsf{n}$_\mathsf{1}$ will be discussed in more detail.

\begin{figure*}[t]
\centering
\includegraphics[width=\columnwidth]{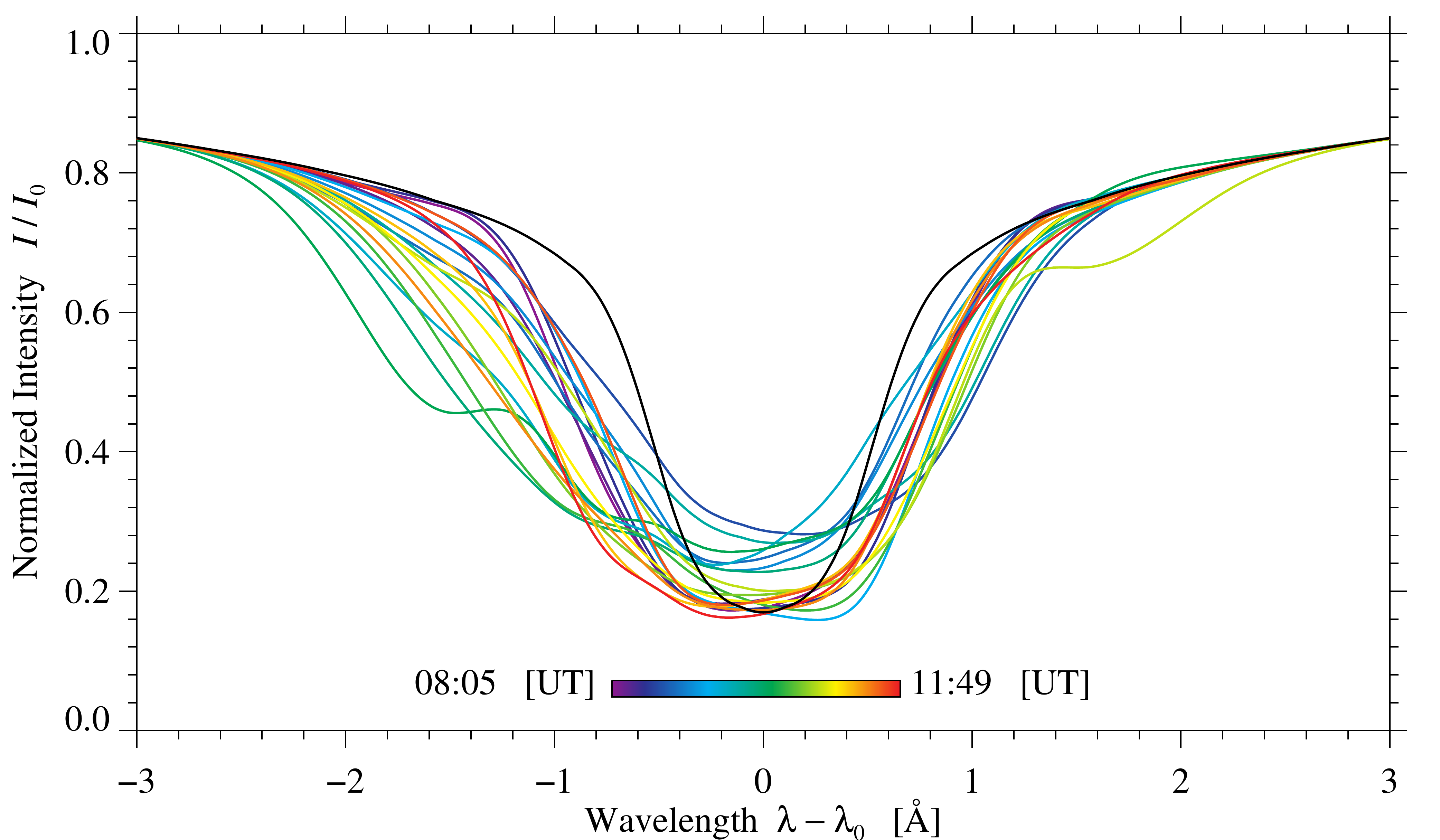}\hfill
\includegraphics[width=\columnwidth]{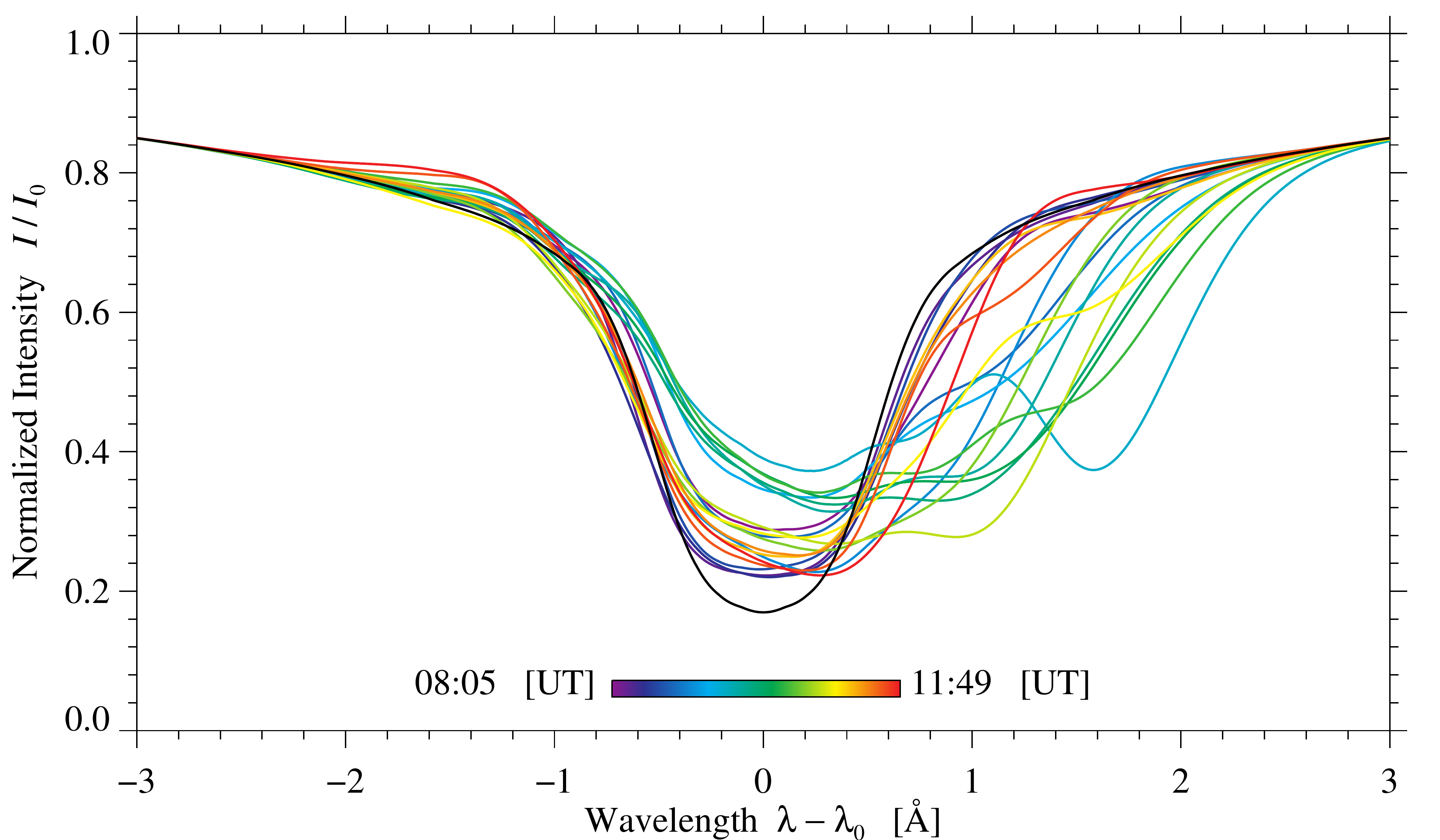}
\caption{H$\alpha$ profiles after PCA decomposition for two regions with
    persistent upflows (\textit{left}) and downflows (\textit{right}) in 
    excess of $\pm$3~km s$^{-1}$ (see online movie with an animated 
    sequence of profiles). The locations are marked in 
    Fig.~\ref{FIG04} by rainbow colored `\textsf{+}' and `$\times$' symbols. 
    The color bar at the bottom uses the same color scheme to track the temporal
    evolution of the spectral line profiles. The black line corresponds to 
    the quiet-Sun profile  in both panels.} 
\label{FIG06}
\end{figure*}

\begin{figure}[t]
\centering
\includegraphics[width=\columnwidth]{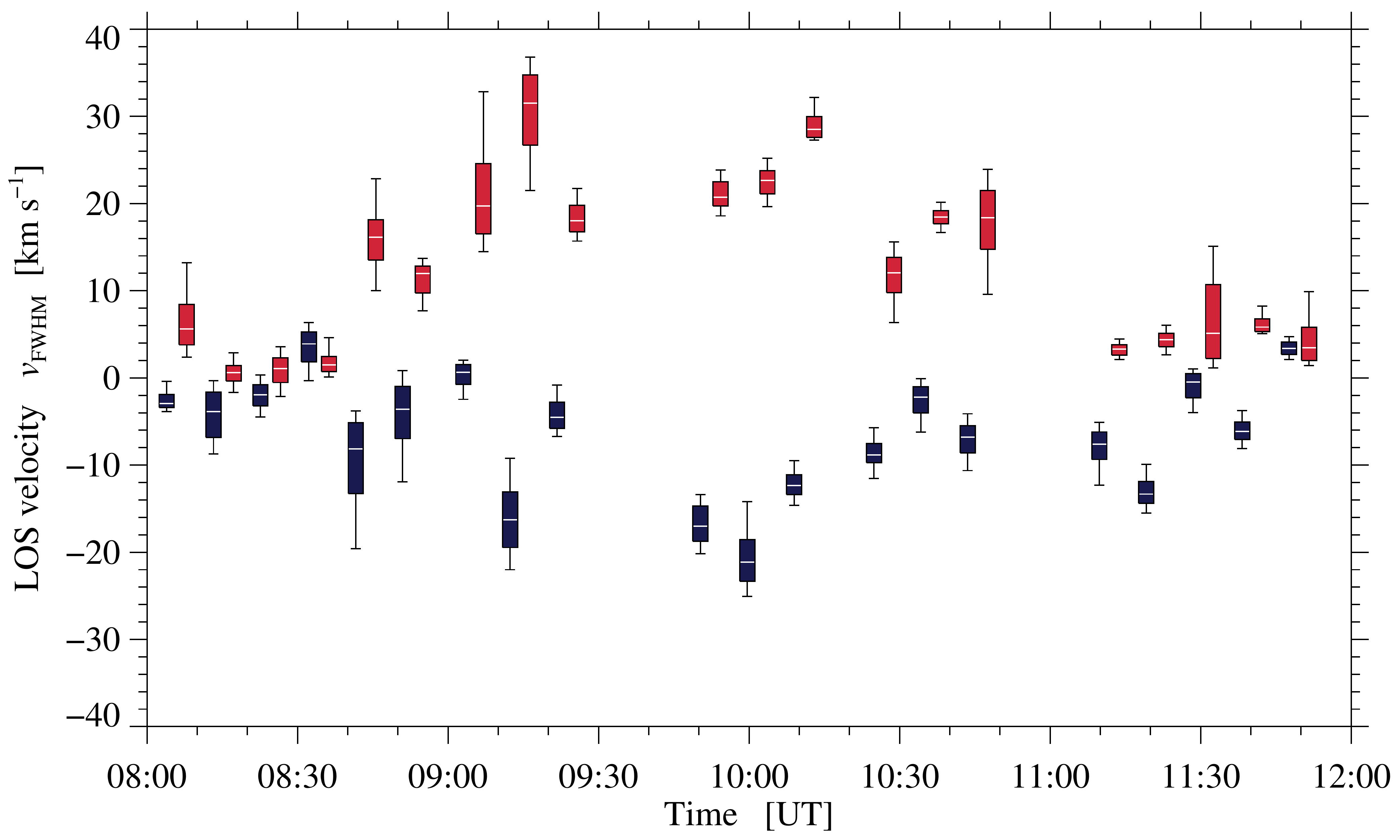}
\caption{Box and whisker plot of the FWHM velocities at the locations that are 
    marked in Fig.~\ref{FIG05} by rainbow colored `\textsf{+}' and `$\times$' 
    symbols. The box extends over one standard deviation above and below the 
    mean velocity of a 10$\times$10-pixel region. The upper and lower whisker 
    refer to the maximum and minimum velocity in that region, respectively. 
    Blue and red refer to the location of persistent up- and downflows,
    respectively. The white horizontal marker corresponds to the median 
    velocity.} 
\label{FIG07}
\end{figure}

Major differences started to show up in the third scan at 08:24~UT leading up to 
the first surge. A fan-shaped region with strong downflows in the center of the 
FOV and a small elongated upflow region became visible in the line-core velocity 
map in the middle of the new flux system, that is, between pore 
\textsf{p}$_\mathsf{1}$ and the negative-polarity feature 
\textsf{n}$_\mathsf{1}$. However, in the FWHM velocity map, the upflow region is 
more prominent and extended, which indicates upward moving plasma that has not 
yet reached the formation height of the line core. Ten minutes later, a small 
fan-like structure forms but with patches of down- and upflows. The upflow 
patches become now conspicuous. The velocity pattern is again changed by the 
fifth scan at 08:43~UT at the location of the first surge. The fan-like 
structure around the mixed polarity region is now replaced by a 
\textsf{V}-shaped feature, where the arms mark locations of downflow with small 
adjacent upflow patches. The pattern of the FWHM velocity is similar to that of 
the line core but downflows are much weaker. 

Strong absorption features (Fig.~\ref{FIG04}) appeared in the scans after 
09:05~UT signalling the initiation of the second surge. In the velocity maps 
(Fig.~\ref{FIG05}), an upflow patch in the line-core velocity map is followed by 
a downflow patch and then by another upflow patch along the green line, which is 
plotted in three panels starting at 09:05~UT. This upflow patch is more 
prominent at the formation height corresponding to the FWHM. At 09:14~UT, the 
upflow patch is more extended and elongated in both the line-core and FWHM 
velocity maps. At this point, the surge reaches its maximum expansion. The 
pattern remains the same for the next scan with a strong upflow patch. For the 
next scans, patches with strong downflows appear. However, in the FWHM velocity 
map, the upflow patch persists and is accompanied by a downflow patch.


\subsection{Evolution in the temporal and spatial domains}\label{SEC35}

The average H$\alpha$ profiles were plotted for all 20 scans in Fig.~\ref{FIG06} 
to illustrate the variety of down- and upflows in the surging region. The black 
profile in both panels refers to the average quiet-Sun profile. Noise was 
significantly reduced in the H$\alpha$ profiles using PCA decomposition, which 
also facilitated the removal of telluric and solar line blends. The location of 
persistent down- and upflows were chosen from the average LOS velocity map and 
by applying a lower threshold of 3~km s$^{-1}$ to the flow speed. Two small 
10$\times$10-pixel regions were selected in the mixed polarity region. In 
addition, these locations are associated with flux cancellation near the 
negative-polarity feature \textsf{n}$_\mathsf{1}$ and the strong proper motion 
of pore \textsf{p}$_\mathsf{2}$ as is evident from the BaSAMs (see 
Fig.~\ref{FIG03}). Downflows are initially located at pore 
\textsf{p}$_\mathsf{2}$, which is responsible for changing the magnetic field 
topology, whereas upflows coincide with the negative-polarity feature 
\textsf{n}$_\mathsf{1}$, which is directly associated to the surge activity. 
These 10$\times$10-pixel regions are marked in Figs.~\ref{FIG03} and \ref{FIG05} 
by `$\times$' and `$+$' symbols referring to locations of persistent down- and 
upflows, respectively. 

\begin{figure*}[t]
\centering
\includegraphics[width=0.33\textwidth]{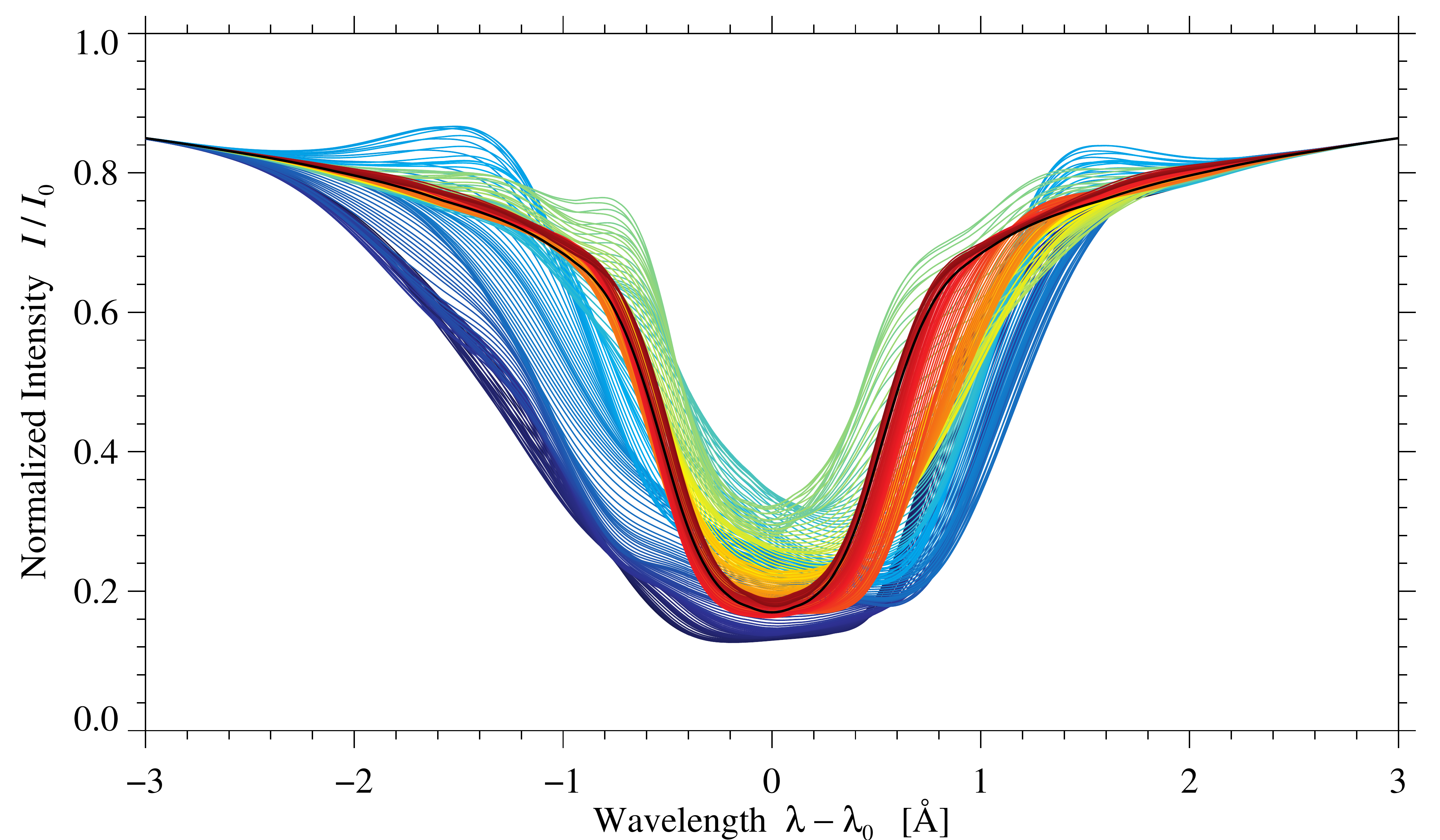}
\includegraphics[width=0.33\textwidth]{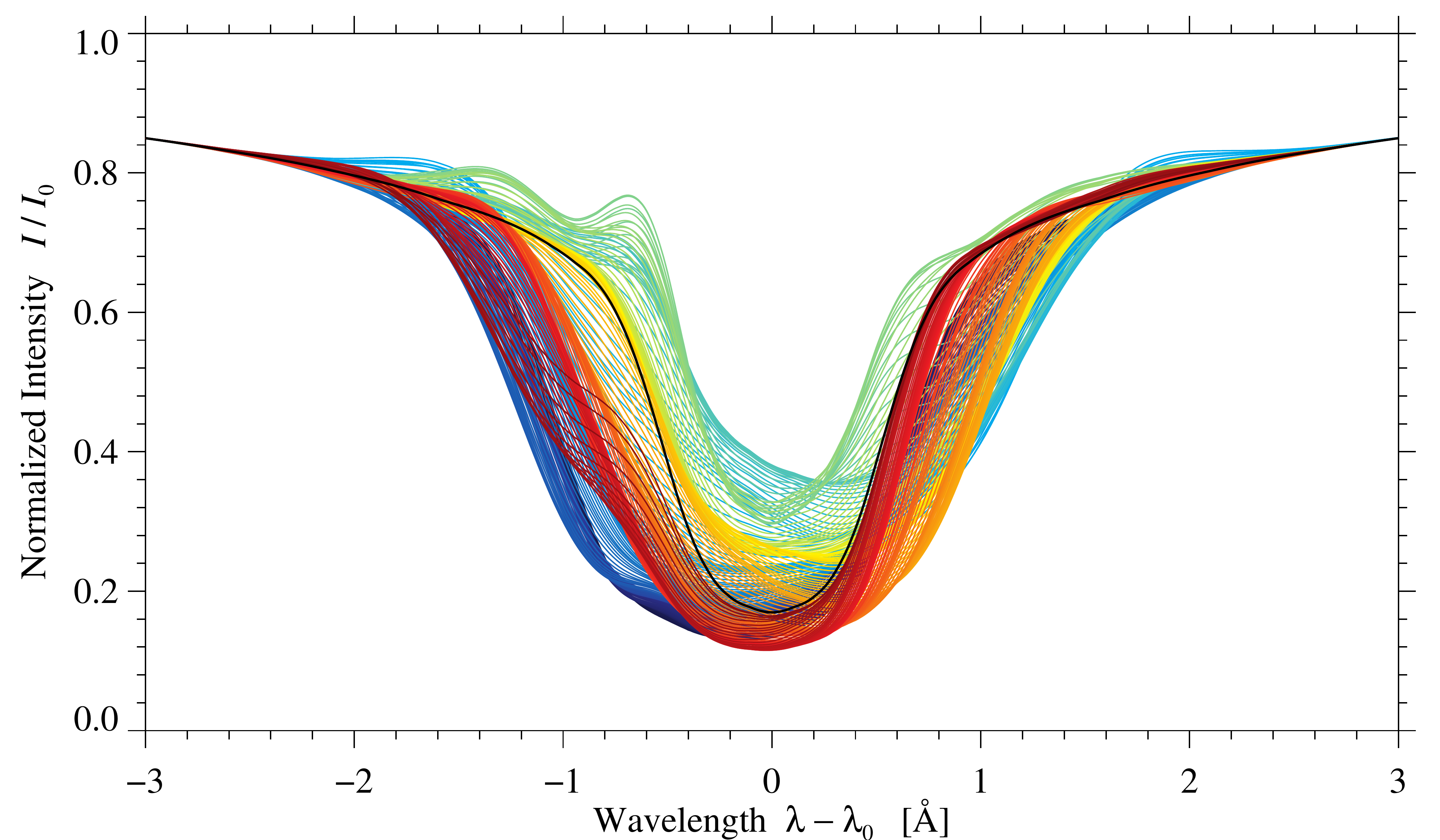}
\includegraphics[width=0.33\textwidth]{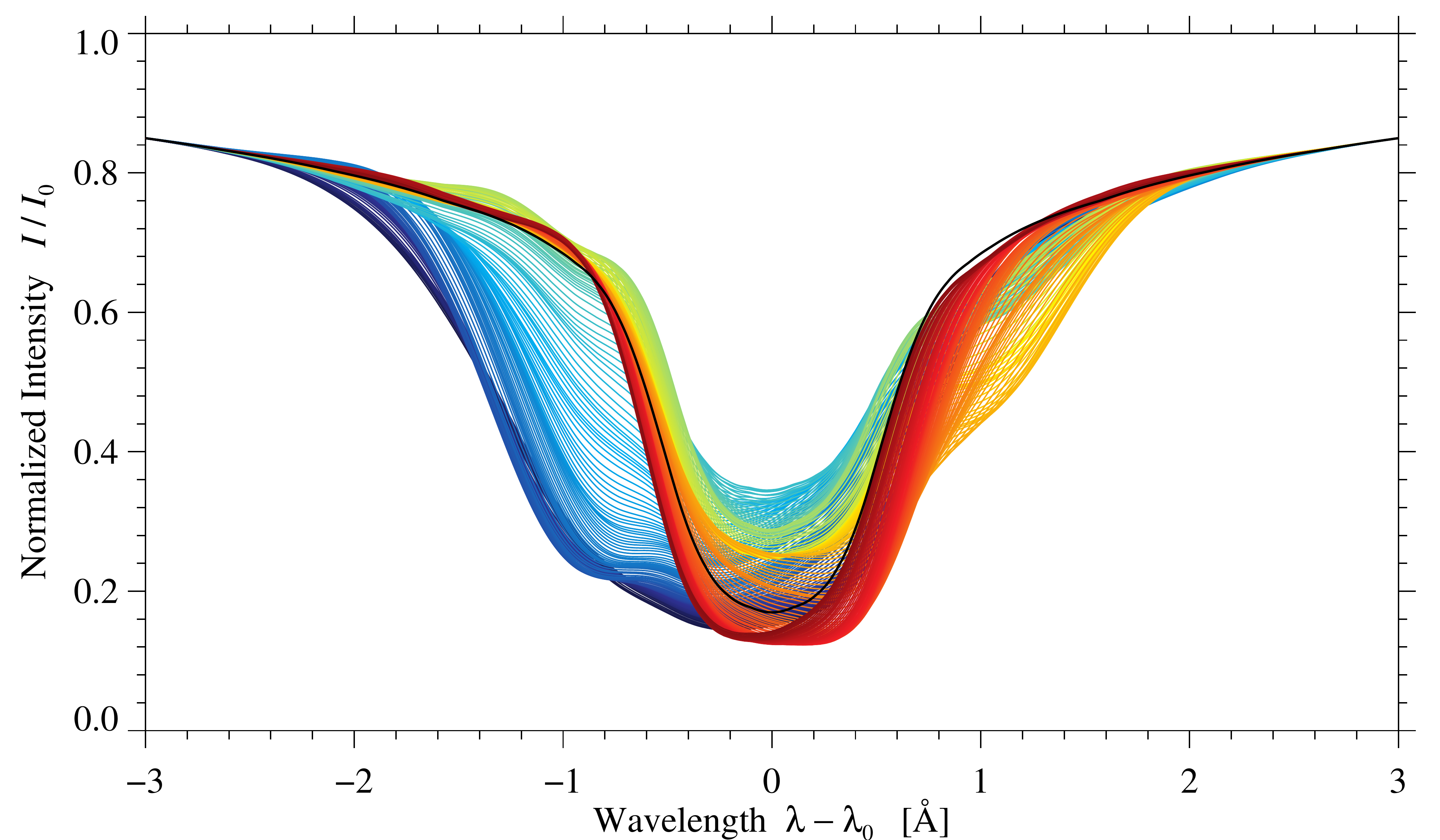}
\includegraphics[width=0.33\textwidth]{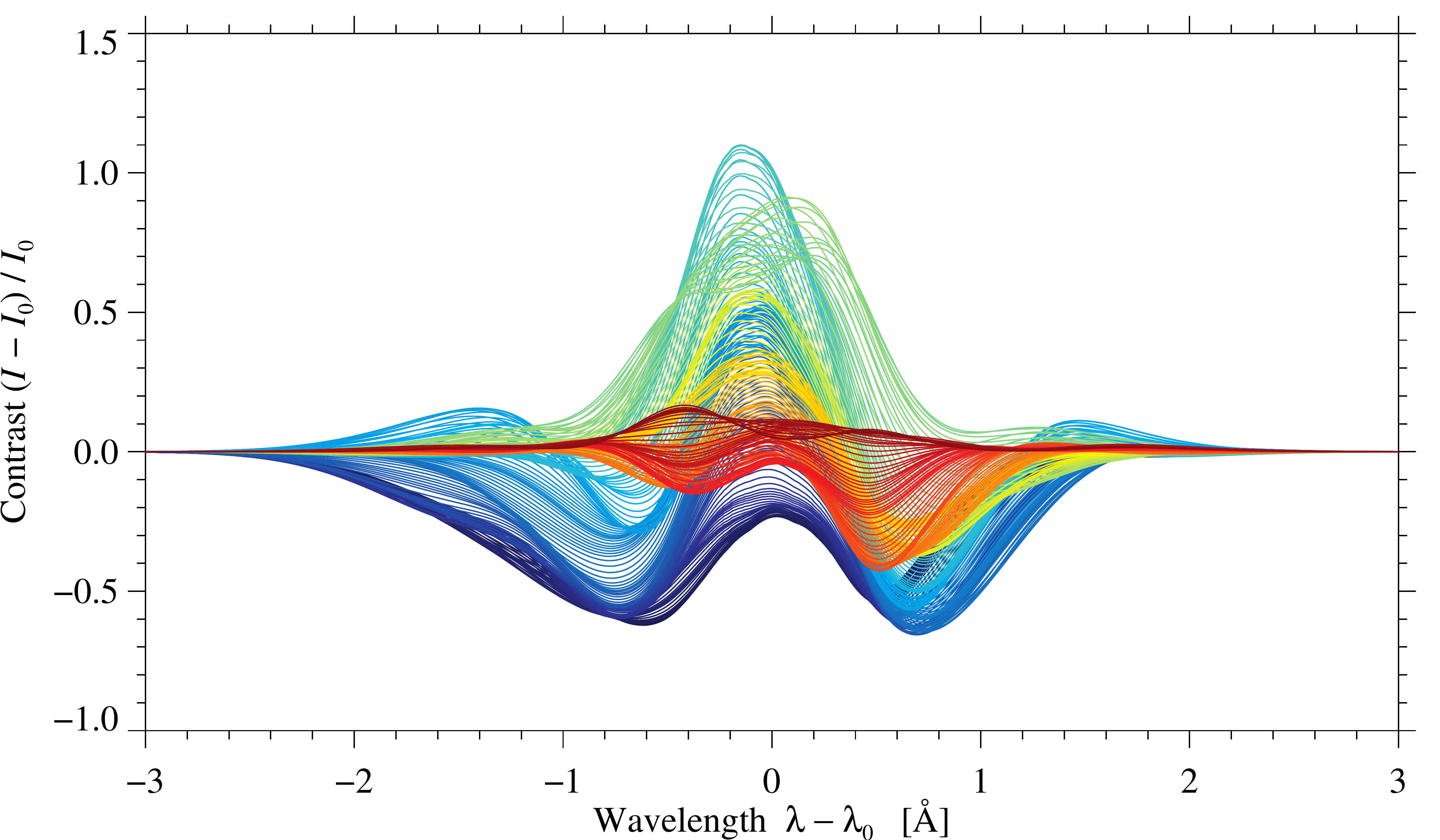}
\includegraphics[width=0.33\textwidth]{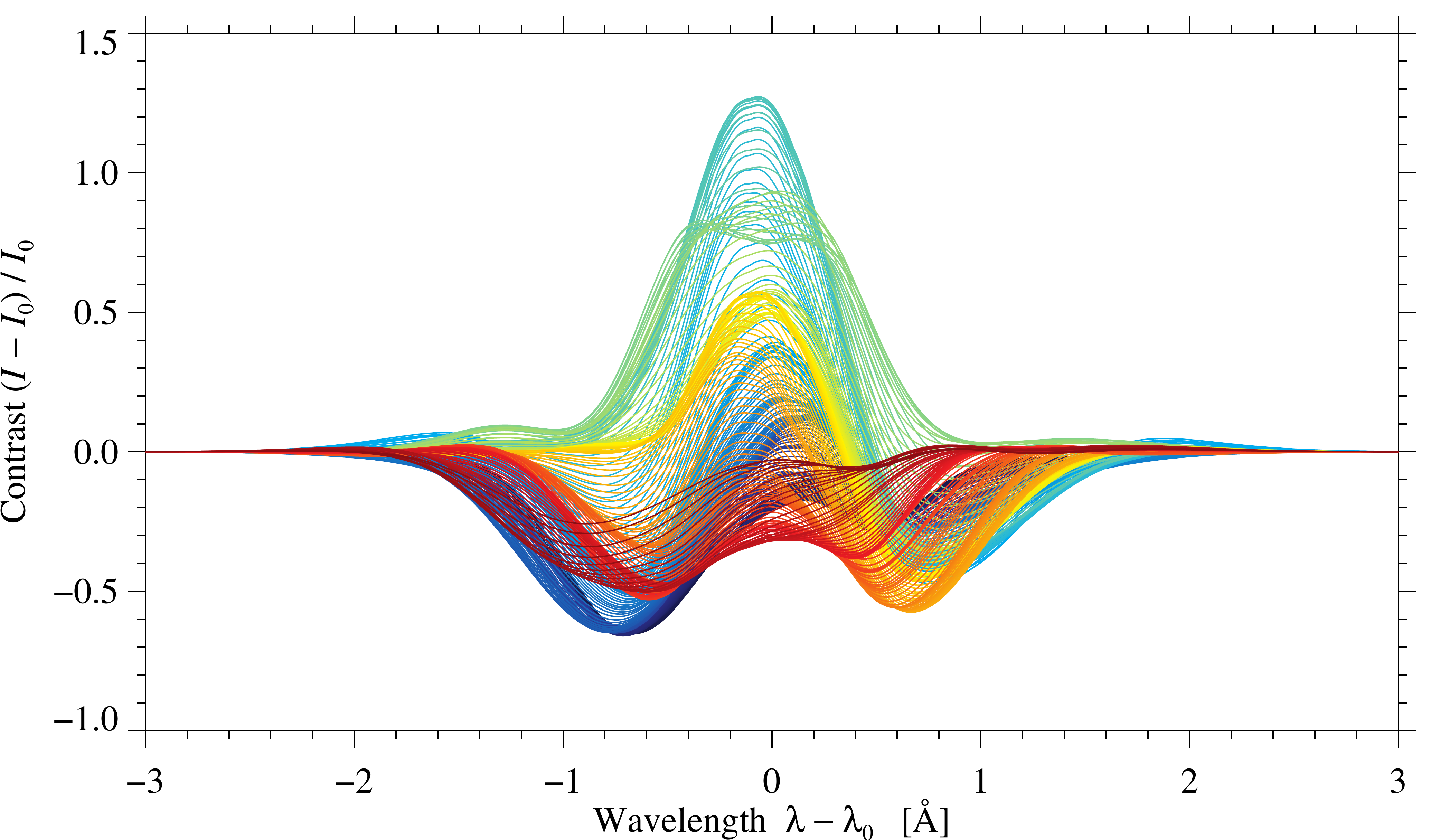}
\includegraphics[width=0.33\textwidth]{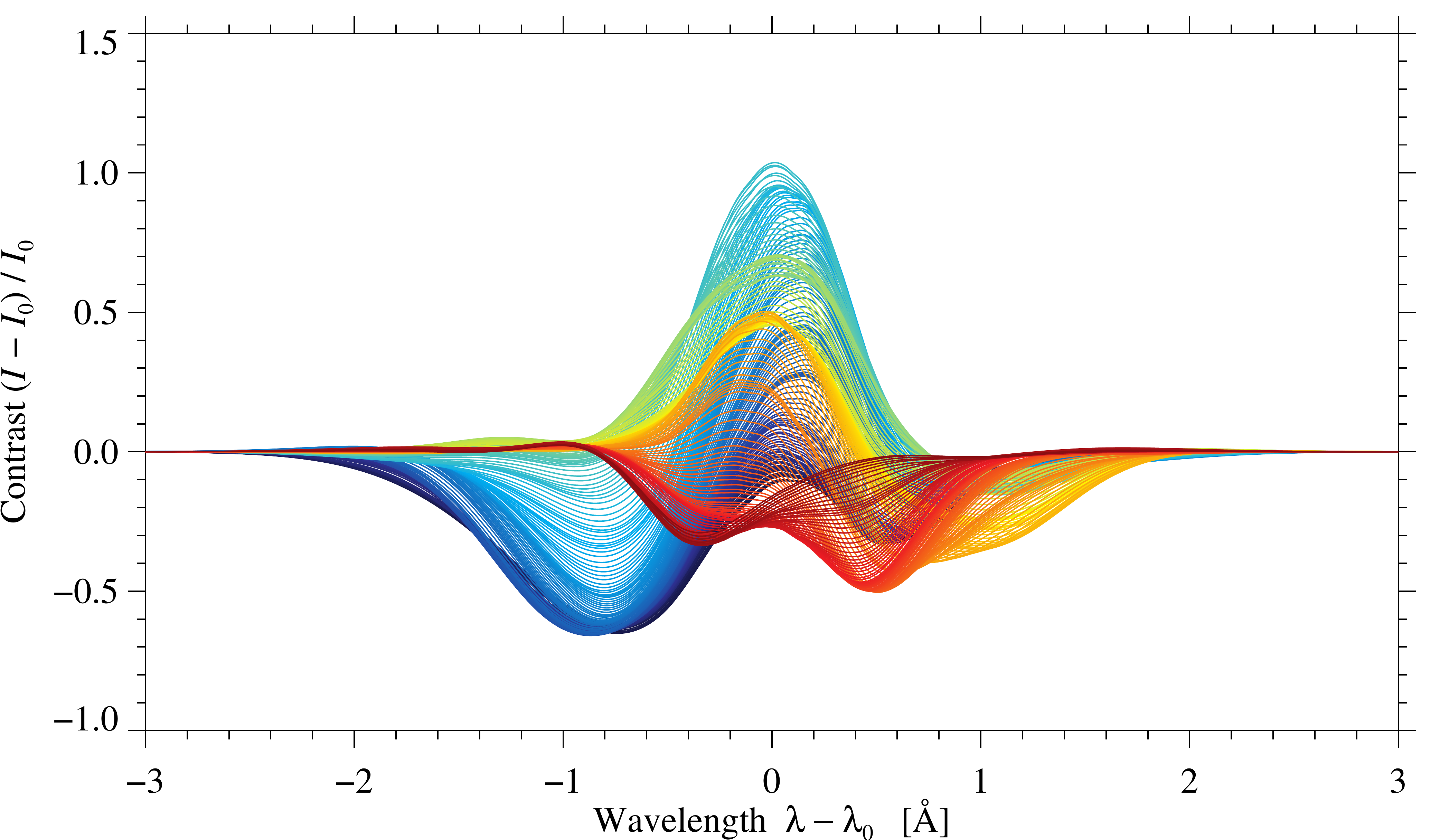}
\caption{H$\alpha$ intensity (\textit{top}) and contrast (\textit{bottom}) 
    along the green cross-section in Fig.~\ref{FIG05}, which starts at the 
    blue and ends at the red square (see online movie with an animated 
    sequence of profiles). Blue and red also mark the start
    and end of the color table for the profiles at 09:05~UT, 09:14~UT, and
    09:23~UT (\textit{left to right}).}
\label{FIG08}
\end{figure*}

\begin{figure}[t]
\centering
\includegraphics[width=0.47\textwidth]{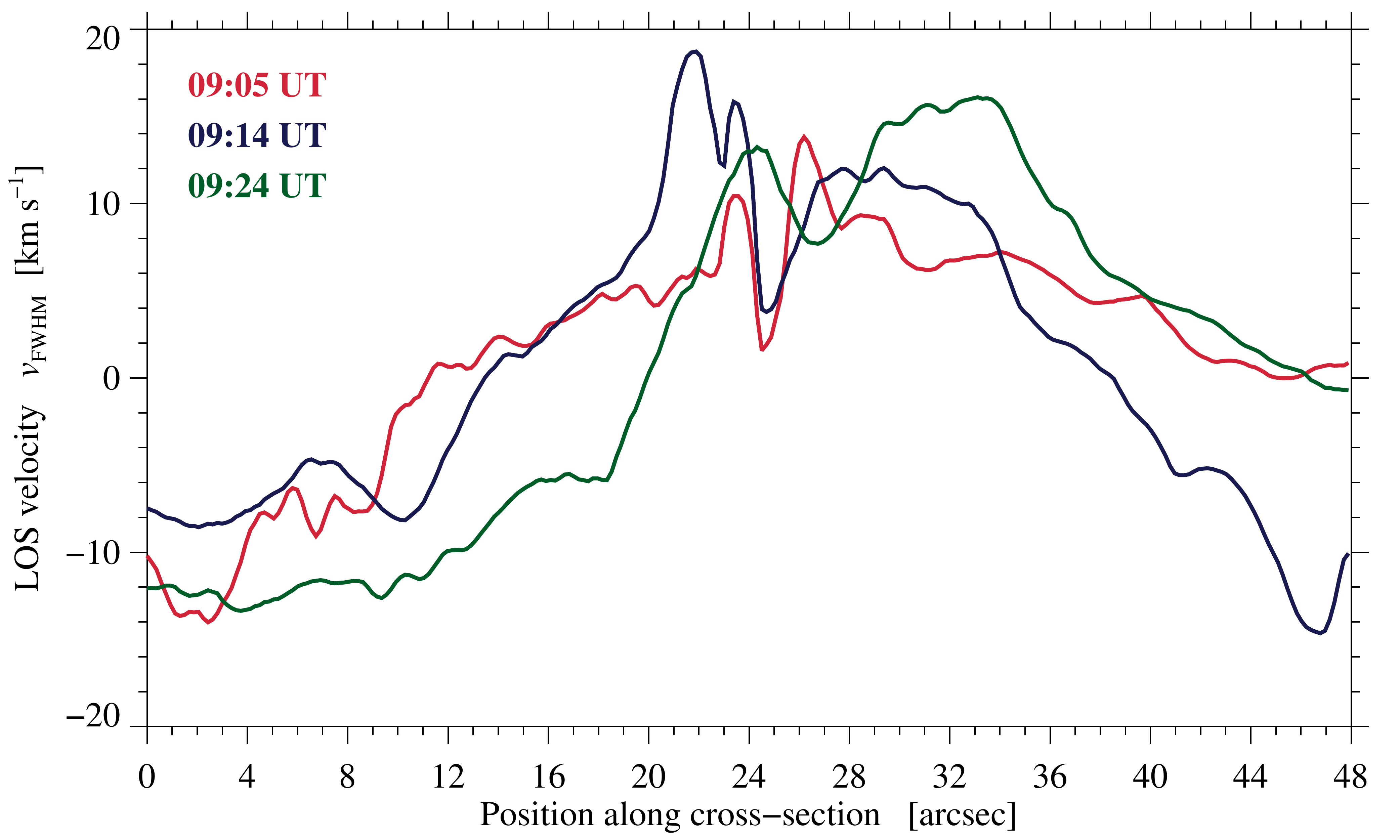}
\caption{FWHM velocities along the three cross-sections in Fig.~\ref{FIG05}.
The zero position corresponds to the blue square and the maximum position to the 
red square of the cross-sections shown as green lines in Fig.~\ref{FIG05}.} 
\label{FIG09}
\end{figure}

Significant changes are seen in the profiles belonging to these two small 
regions. Blueshifted profiles (left panel in Fig.~\ref{FIG06}) are slightly 
displaced for the first two scans but much less in comparison to shifts for 
later scans. The line core is deeper compared to an average quiet-Sun line 
profile. Starting with the third scan, the profiles become shallower and 
line-core intensities reach almost $I / I_0 = 0.3$ at the onset of the first 
surge. In the initial phase of the second surge from 09:05\,--\,09:23~UT, the 
blue-green profiles are shifted up to $-1$~\AA\ implying Doppler velocities of 
about 45~km s$^{-1}$. Double-lobed profiles and profiles with extended shoulders 
indicate more complex plasma motions either because of features with different 
Doppler velocities in the same resolution element or because of height-dependent 
flows. Typically, fast and slow flow components are present in double-lobed 
profiles. During the time period 09:52\,--\,10:10~UT (green) in the later stages 
of the second surge, the blueshifts reached up to $-2$~\AA\ with corresponding 
plasma velocities of about \mbox{90~km s$^{-1}$}. Thus, the second surge was 
most clearly visible for a period of about one hour after 09:05~UT. After this 
the blueshifted profiles slowly reach quiet-Sun line-core intensities but remain 
broadened, which indicates turbulent and/or heated plasma. The H$\alpha$ line is 
an important diagnostic to probe the solar chromosphere \citep{Carlsson2019}. 
However, the low atomic weight of Hydrogen leads to large thermal broadening. 
Hence, the changes in the H$\alpha$ line widths are indicative of the 
temperature changes in the deep photosphere \citep{Cauzzi2009}, whereas changes 
in line-core intensity indicate variations of the formation height 
\citep{Leenaarts2012} because of strong scattering and the sensitivity of 
irradiation from the surrounding environment.

Redshifted profiles follow the same trend as blueshifted profiles. However, they 
are always shallower and never reach the quiet-Sun line-core intensity. 
Furthermore, profiles in the later stages of evolution exhibit enhanced line 
wings. Line profiles are narrow in the first scan at 08:05~UT but this changes 
rapidly in the next scans with the onset of the first surge, when the line 
profiles become broader and reveal a distinct shoulder in the red wing, which 
extends to +1~\AA\ and beyond. The profiles during the intervening time interval 
from 08:43\,--\,09:05~UT, that is, between the two surges, are very broad and 
shallow. The shoulder reaches almost 2~\AA\ for the scans at 09:14~UT and 
09:23~UT once the second surge started. These profiles even show a clearly 
separated second lobe and appear at the same time in proximity to the surge. The 
initially redshifted profiles also return to quiet-Sun conditions towards the 
end of the observations. The changes in individual profiles can be visualized in 
the animated online version of Fig.~\ref{FIG06}.

To quantify the morphological diversity and the trend in the temporal evolution 
of spectral line profiles, a box and whisker plot (Fig.~\ref{FIG07}) is created 
of the FWHM velocities for the locations marked in Fig.~\ref{FIG05}. The FWHM is 
chosen to take into consideration the complete line (asymmetric) profile rather 
than just the line core, which may not be representative for strongly asymmetric 
profiles. In general, velocities for the satellite component of double-lobed 
profiles cannot be appropriately derived by bisector analysis. The 
10$\times$10-pixel regions are used to compute statistical properties such as 
the mean, standard deviation, median, maximum, and minimum velocity values, 
which are depicted as center and height of the box, white horizontal marker, 
upper whisker and lower whisker, respectively. In the early scans at 
08:05\,--\,08:34~UT, velocities at locations of persistent up- and downflows are 
insignificant and show small variation. Starting at 08:43~UT, the speeds at both 
locations start to rise, reaching a maximum of up to $\pm$20~km~s$^{-1}$. The 
up- and downflow velocities are synchronized and consistently increase until 
10:01~UT.

Three cross-sections along the second surge reveal the spatial variation of the 
H$\alpha$ intensity and contrast profiles at 09:05~UT, 09:14~UT, and 09:23~UT, 
respectively. The displayed contrast profiles refer to $(I - I_{0})/I_{0}$, 
where $I_{0}$ is the quiet-Sun intensity. The extracted profiles are collected 
in Fig.~\ref{FIG08}, which contains 256 profiles in each panel along the 
cross-sections shown in Fig.~\ref{FIG05}. The profiles in Fig.~\ref{FIG08} 
follow a color scheme from blue to red, which indicates their location along the 
cross-section. The ends of the cross-section are marked by blue and red squares 
in Fig.~\ref{FIG05}. The intensity and contrast profiles vary significantly 
along the cross-section. The three scans cover three different stages of the 
surge. The second surge appears at the bottom of the cross-section at 09:05~UT 
and expands within 20~min to the south. The upper end of the cross-section 
extends towards a quiet-Sun region, whereas the middle part passes through a 
redshifted region.

The intensity and contrast profiles are significantly broadened and blueshifted 
at 09:05~UT at the start of the cross-section, as indicated by the dark to light 
blue profiles in the left panel of Fig.~\ref{FIG08}. Their line-core intensities 
are initially lower compared to a quiet-Sun profile but the profiles become 
rapidly shallower, as is clearly evident from the contrast profiles. These 
profiles belong to the tip of the surge. Hence, profiles corresponding to strong 
upflows are expected. Profiles from the central part of the cross-section are 
still broad but they gradually exhibit a second, redshifted component. Towards 
the base of the surge, the redshifts increase, signaling strong downflows, and 
the profiles show enhanced line-wing emissions. The orange and red profiles at 
the end of the cross-section are slightly broadened and redshifted but in 
general, they resemble quiet-Sun profiles. Indeed, the profiles closest to the 
end of the cross-section are very good representations of the quiet-Sun. The 
discussed properties of intensity profiles are reflected in the contrast 
profiles, which clearly accentuate line depth and width as well as line 
asymmetries and Doppler shifts.

For the scan at 09:14~UT, the lower tip of the surge still exhibits blueshifts 
but with narrower line profiles. The slight line-wing enhancement is still 
present in the turquoise profiles. Further along the cross-section, redshifted 
spectral lines are encountered as indicated by the yellow-orange profiles. A 
blueshifted absorption region is located at the upper end of the cross-section, 
which is also reflected in the red profiles with two components. The sequence of 
profiles changes more gradually along the cross-section at 09:23~UT. The blue 
profiles are clearly blueshifted and become redshifted when progressing along 
the cross-section. A line-wing enhancement is not evident for the third 
cross-section. However, some of the blue profiles show a hint of a second 
component. For better visualization animated version of the Fig.~\ref{FIG08} is 
available online.

slit 
\begin{figure}[t]
\centering
\includegraphics[width=0.98\columnwidth]{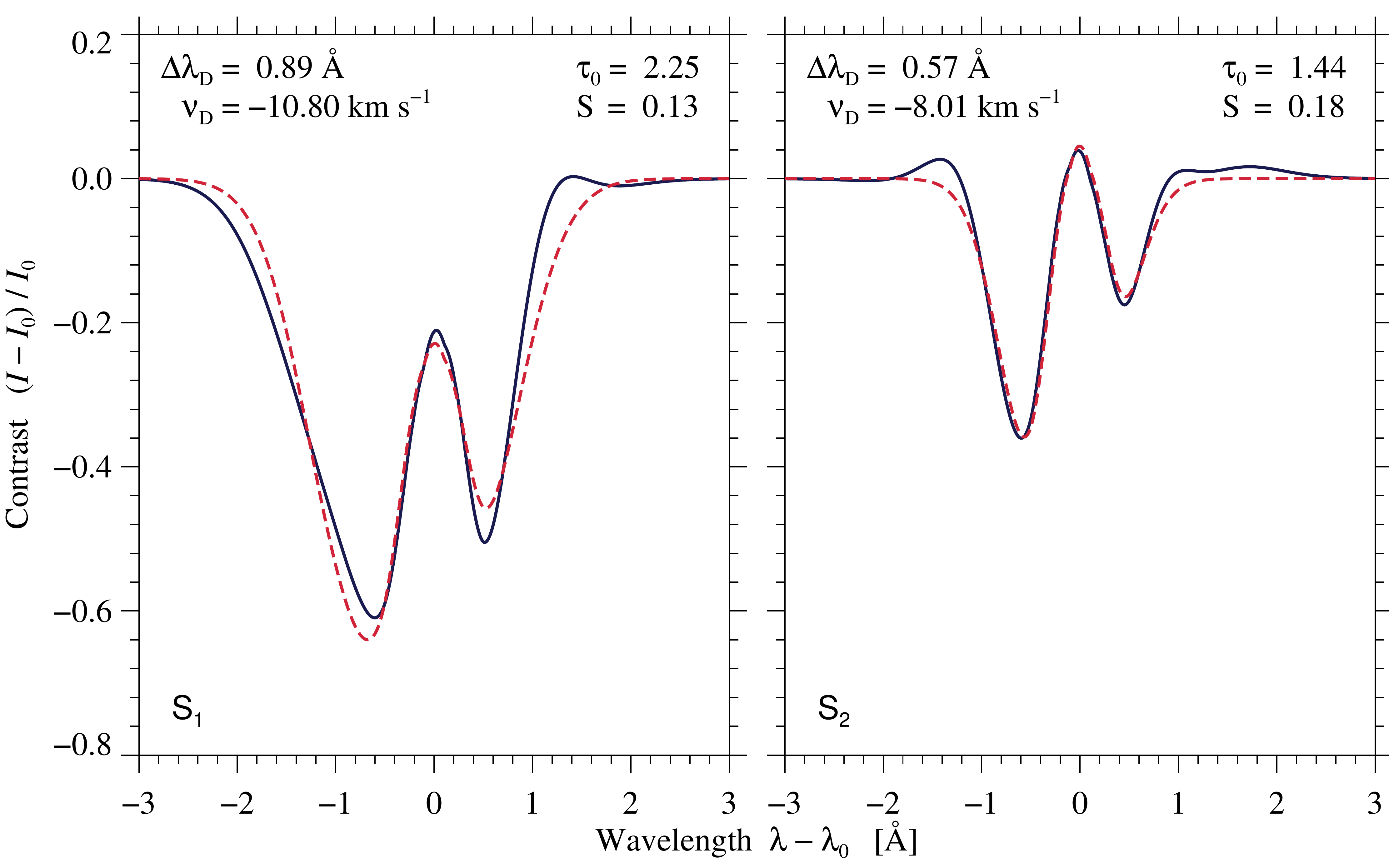}
\caption{PCA decomposed (\textit{solid blue}) and CM (\textit{dashed red})
    H$\alpha$ contrast profiles for two locations \textsf{S}$_\mathsf{1}$ 
    (\textit{left}) and \textsf{S}$_\mathsf{2}$ (\textit{right}) within the 
    surge (see Fig.~\ref{FIG04}). The CM parameters are given at the top of 
    the panels.} 
\label{FIG10}
\end{figure}

Figure~\ref{FIG09} summarizes and quantifies the variation of the FWHM 
velocities along the three cross-sections in Fig.~\ref{FIG05}. The overall 
variation along the cross-sections is similar but with a distinct shift for the 
third cross-section. Strong upflows between $-8$ and $-12$~km~s$^{-1}$ are 
encountered to the south of the base of the second surge near the 
negative-polarity feature \textsf{n}$_\mathsf{1}$ while the cross-section comes 
across downflows of up to 20~km~s$^{-1}$ near the rapidly evolving and migrating 
pore \textsf{p}$_\mathsf{2}$. The end of the cross-section is located in a 
quiet-Sun region.


\subsection{Cloud model inversions}\label{SEC36}

Properties of spectral absorption lines are often more conveniently determined 
using the contrast profiles. In Fig.~\ref{FIG08}, contrast profiles along the 
three cross-sections are presented and CM inversions are performed for all 20 
scans as mentioned in Sect.~\ref{SEC02}. The details of CM inversions based on 
PCA noise-stripped contrast profiles are explained in \citet{Dineva2020}. To 
illustrate the quality of the CM inversions, two examples of PCA decomposed and 
CM H$\alpha$ contrast profiles are shown in Fig.~\ref{FIG10}. The profiles are 
taken from the two positions in the cross-section at 09:05~UT at the onset of 
the second surge and are marked as \textsf{S$_1$} and \textsf{S$_2$} in the 
line-core map of Fig.~\ref{FIG04}. Both profiles belong to dark absorption 
features. However, the optical thickness is higher for \textsf{S$_1$} with 
$\tau_{0} = 2.25$ than for \textsf{S$_2$} with $\tau_{0} = 1.44$. The cloud 
velocity is also higher for \textsf{S$_1$} with $v_{D} = -10.80$~km~s$^{-1}$ 
than for \textsf{S$_2$} with $v_{D} = -8.01$~km~s$^{-1}$. In addition, 
significant line broadening is evident for \textsf{S$_1$} with 
$\Delta\lambda_{D} = 0.89$~\AA\ as compared to $\Delta\lambda_{D} = 0.57$~\AA\ 
for \textsf{S$_2$}. Broadened line profiles, strong absorption, and strong 
upflows thus characterize the site where the surge originates.

\begin{figure*}[t]
\centering
\includegraphics[width=\textwidth]{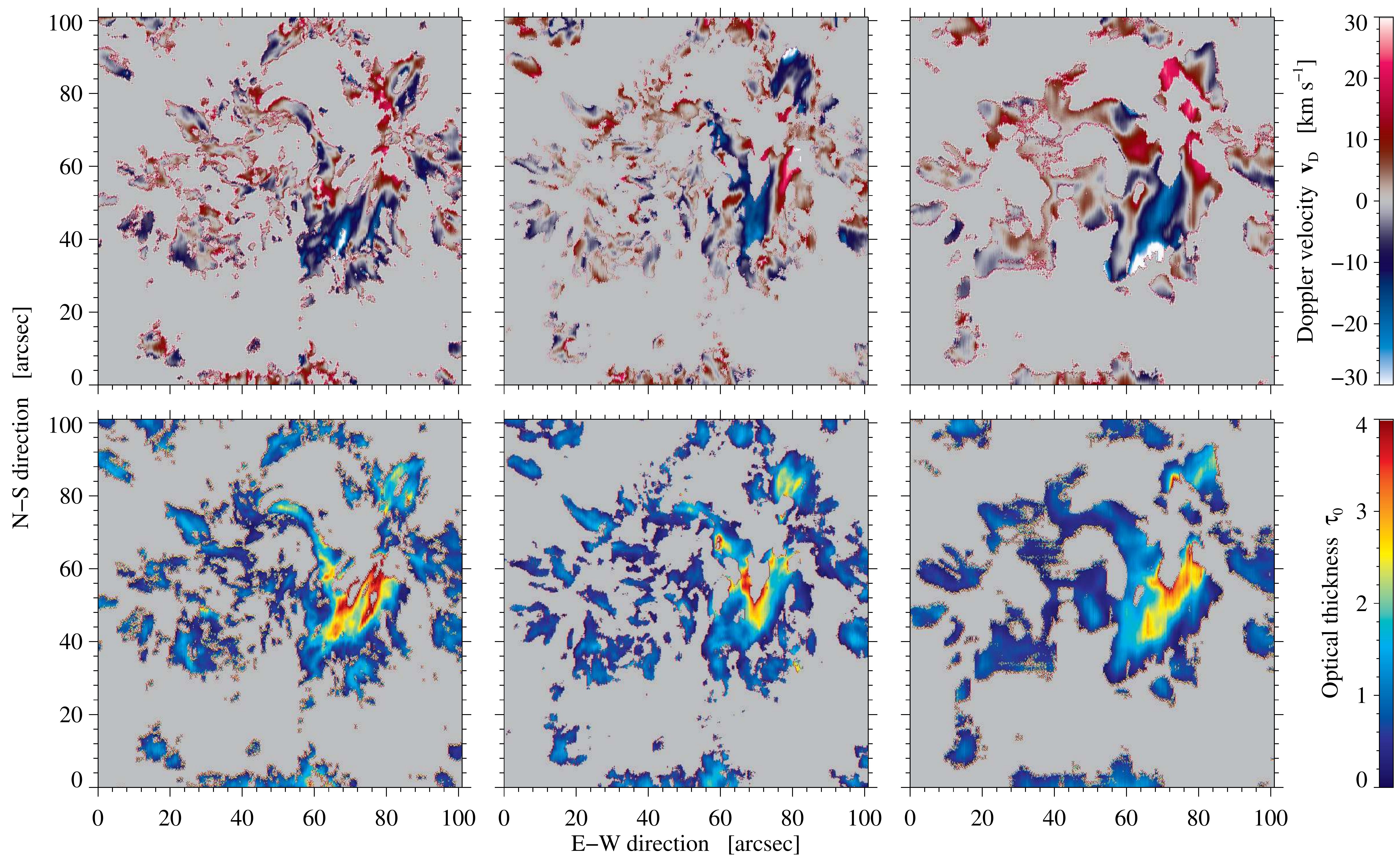}
\caption{Two-dimensional maps of the inversion results for the two CM 
    parameters based on the noise-stripped contrast profiles for three scans at 
    09:05~UT, 09:14~UT, and 09:23~UT (\textit{left to right}): Doppler velocity 
    of the cloud material $v_\mathrm{D}$ (\textit{top}) and optical thickness
    $\tau_0$ (\textit{bottom}). The FOV is the same as in Figs.~\ref{FIG04} 
    and \ref{FIG05}. Note that the velocity of the cloud material 
    $v_\mathrm{D}$ differs from those depicted in Fig.~\ref{FIG05}. 
    Regions, where the CM is unsuitable and inversions fail, are reproduced
    in gray.} 
\label{FIG11}
\end{figure*}

Two-dimensional maps of the Doppler velocity of the cloud material 
$v_\mathrm{D}$ and the optical thickness $\tau_0$ are depicted in 
Fig.~\ref{FIG11} for three scans at 09:05~UT, 09:14~UT, and 09:23~UT. Note the 
deteriorating seeing in the last scan resulting in a washed-out appearance. The 
three selected scans cover the second surge from onset to roughly its peak and 
complement the exploration of the three cross-sections in Fig.~\ref{FIG05}. The 
cloud velocities are significantly higher than the ones shown in 
Fig.~\ref{FIG09}. The CM is based on the assumption that cool absorbing plasma 
is suspended at chromospheric heights by the magnetic field and that cloud 
material is irradiated by the photosphere from below and with contributions from 
the surroundings. The upflows are highly structured and are later accompanied by 
downflows in close proximity. These simultaneous blue- and redshifts were also 
noted by \citet{Tiwari2019}, however, in IRIS UV spectra. The side-by-side up- 
and downflow of plasma along the surge argues according to these authors against 
twisting motions in the surge. The upflows are directed towards the south but do 
not reach the southern edge of the FOV. The highest plasma speeds of more than 
30~km~s$^{-1}$ are encountered at the leading edge of the surge and are highest 
at 09:23~UT.

The optical thickness $\tau_0$ is very high at the base of the surge. The high 
optical thickness in proximity to the base of the surge was also observed by 
\citet{Schmieder1994}. CM inversions, however, fail directly at the base of the 
surge because of enhanced line-core intensity, which signals local heating of 
the plasma. This is also supported by the larger Doppler width 
$\Delta\lambda_\mathrm{D}$ at these locations. The optical thickness decreases 
in the propagation direction of the surge. At some point, the cool plasma 
becomes either so diluted that only the quiet-Sun photosphere is observed or the 
ejected plasma is heated to temperatures where H$\alpha$ becomes ionized. This 
relatively sharp transition finds its counterpart in the SDO HMI/AIA online 
movie accompanying Fig.~\ref{FIG01}, that is, in the EUV $\lambda$304~\AA\ 
time-series, the initially opaque dense plasma brightens and can be traced to 
the edge of the FOV (see Sect.~\ref{SEC32}).

In addition to the surge, a stable arch filament with upflows is present in the 
left panels of Fig.~\ref{FIG11} above the center of the FOV. This points to the 
complex magnetic structure of the active region. However, it also indicates the 
co-existence of the old and new flux systems, which both experience continuous 
flux emergence. Only in the EUV $\lambda$304~\AA\ time-series enhanced 
brightenings appear in the loop system associated with the arch filament (see 
Sect.~\ref{SEC32}).


\section{Discussion}\label{SEC04}

Surges occur at sites of flux cancellation, that is, in mixed-polarity regions 
of an EFR, when trailing and leading parts of the active region separate 
\citep{Wang1992}. Surges were also associated with the cancellation of 
opposite-polarity MMFs around sunspots at the periphery of the superpenumbra 
\citep{Wang1991b}. In both cases, surges feed of the energy supplied by magnetic 
cancellation and are often accompanied by \mbox{(sub-)flares}, which are, 
however, absent in the present observations. In the present study, flux 
emergence between the pores \textsf{P}$_\mathsf{2}$ and \textsf{p}$_\mathsf{1}$ 
continues throughout the observations, which is accompanied by strong proper 
motions of pore \textsf{p}$_\mathsf{2}$ affecting the magnetic field topology of 
the active region. Additionally, prolonged flux cancellation occurs near the 
negative-polarity feature \textsf{n}$_\mathsf{1}$, which defines the base of the 
surge.

\citet{Zirin1967} reported homologous surges, that is, recurring surges with 
similar shape and evolution \citep[see also][]{Chen2008}. In the present study, 
two homologous surges were detected during the four-hour observing run, which 
can be clearly identified when using H$\alpha$ observations in combination with 
SDO EUV images. \citet{Nelson2019} suggests that multiple ejections with similar 
characteristics hint at repetitive drivers, for example, reconnection in the 
lower atmosphere. However, in the present study, flux cancellation at the 
negative-polarity feature \textsf{n}$_\mathsf{1}$ is rather continuous than 
repetitive, relaxing the strict requirement of recurrent reconnection. The 
heating at the base of the surges is noted as continuous brightenings in 
H$\alpha$ line-core maps, broad line profiles, and in time-sequence of UV and 
EUV images as localized brightenings. This is consistent with the established 
concept that heating in the lower atmosphere pushes plasma higher in the 
atmosphere during the surge \citep[e.g.,][]{Schmieder1994}. Two-dimensional 
magneto-hydrodynamics simulations by \citet{Yang2018} of a network jet triggered 
by magnetic reconnection in the transition region during flux emergence manifest 
that relatively cool jets are followed by hot fast jets, as observed for surges. 
The returning plasma of the observed surge does not carry sufficient momentum to 
disturb the surrounding chromosphere, as was observed by \citet{Zirin1967}, 
where the impact creates a travelling wave, which in turn caused other surges or 
flares. In contrast, a regular AFS co-exists undisturbed during the homologous 
surges. However, the broad H$\alpha$ spectral profiles near the base of the 
surge suggest heated and turbulent plasma even in the deep photosphere.

In general, the observed FWHM velocities of about 20~km~s$^{-1}$ are well above 
the sound speed of about 10~km~s$^{-1}$ but remain at the lower end of values 
reported in literature, for example, about 30~km~s$^{-1}$ by \citet{Chen2008}, 
40~km~s$^{-1}$ by \citet{Watanabe2011}, 70~km~s$^{-1}$ by \citet{Yang2014}, 
100~km~s$^{-1}$ by \citet{SanchezAndradeNuno2008}, and 200~km~s$^{-1}$ by 
\citet{Bruzek1974}. Even the CM velocities of more than 30~km~s$^{-1}$ remain at 
the lower end of the cited studies and stay well below the Alfv\'en speed of 
about 100~km~s$^{-1}$. Only some of the double-lobed H$\alpha$ line profiles 
imply plasma motion reaching the Alfv\'en speed. In any case, some ambiguity 
remains depending on the methods that were used to determine the surge 
velocities, that is, time-space slices, feature tracking, line-core fitting, 
bisector analysis, CM inversions, or modeling of dual- or multi-component 
spectral line profiles. 

In addition, the complexity of spectral line profiles may not be appropriately 
captured by imaging spectrometers with lower spectral resolution, for example, 
only five spectral points were covered in H$\alpha$ by \citet{Madjarska2009} and 
23 in \citet{Watanabe2011}. Even in recent studies of H$\alpha$ surge 
observation this limitation remains, for example, \citet{Yang2019} used nine 
wavelength points, the data of \citet{Nelson2019} had 35 wavelength points, and 
\citet{Ortiz2020} used 15 wavelengths points. Currently, spectroscopic H$\alpha$ 
observations are only available from the ground. Thus, even in the presence of 
real-time adaptive optics correction and image restoration, variable seeing 
conditions will impact the analysis of spectral profiles. In spectral 
inversions, these impacts may even be larger than the intrinsic error of the 
inversion techniques. In the meantime, the Chinese Solar H$\alpha$ Imaging 
Spectrometer \citep[SHIS,][]{Chen2018} spacecraft recently received approval and 
in the future, H$\alpha$ spectral data free of seeing influence will be 
available. 

Surges are a common phenomenon in EFRs and sites of flux cancellation. However, 
investigations of the ejected surge plasma are not frequently carried out using 
high-spectral resolution H$\alpha$ spectroscopy. The present investigation 
combines moderate temporal (about 10-minute cadence) with good spectral 
(4.2~m\AA\ pixel$^{-1}$ dispersion) and good spatial (better than one second of 
arc) resolution, which allowed us to trace the detailed temporal evolution and 
spatial variation of a surge. The spatio-temporal properties of the H$\alpha$ 
line profiles in Figs.~\ref{FIG06} and~\ref{FIG08} clearly demonstrate that the 
velocity distribution along the surge is highly structured and very variable in 
time. Higher spatial resolution can be achieved with imaging spectrometers such 
as CRISP \citep[e.g.,][]{Scharmer2008b, Watanabe2011} but with much lower 
spectral resolution. On the other hand, the Fast Imaging Solar Spectrograph 
\citep[FISS,][]{Chae2013} delivers resolved spectral observations in H$\alpha$ 
and Ca\,\textsc{ii} $\lambda$8542~\AA\ and provides higher spatial resolution 
but over a smaller FOV compared to the VTT echelle spectrograph.


\section{Conclusions}\label{SEC05}

The aim of the present study was to demonstrate the potential of high-resolution 
spectroscopy for the investigation of solar eruptive phenomena, that is, two 
homologous surges in the present case. Resolved line profiles are rare and often 
spectral features are obscured by resorting to filtergrams and narrow-band 
images. Furthermore, noise in addition to solar and telluric blends affects the 
analysis of strong chromospheric absorption lines. Improvements in data 
processing \citep[see][]{Dineva2020} addressed these issues, and this is the 
first scientific investigation that benefits from these software tools 
implemented in the ``sTools'' IDL software library \citep{Kuckein2017}. 

The high-resolution noise-stripped H$\alpha$ observations in conjunction with 
SDO LOS magnetograms and UV/EUV images revealed various physical properties of 
the observed surges. The homologous surges appear in a region of new flux 
emergence. The interaction of new emerging flux with pre-existing magnetic flux 
concentrations provides suitable conditions, that is, flux cancellation and 
strong proper motions required for the seen surge activity. The two events of 
enhanced surge activity are related to the locations of flux cancellation and 
strong proper motions, which in turn are the locations of persistent 
side-by-side down- and upflows. Furthermore, the surges associated with the new 
emerging magnetic flux co-exist with the stable AFS established within the 
preexisting active region magnetic flux. Broad and dual-lobed H$\alpha$ profiles 
are prevalent along the surge, representing accelerated/decelerated flows of 
cool plasma with complex structure. Additionally, broad H$\alpha$ profiles 
indicate enhanced heating at the base of the surge, that is, strong gradients in 
temperature and density, where cool plasma in H$\alpha$ absorption features is 
located above heated regions. This scenario is also supported by the presence of 
bright kernels in UV/EUV images. The opacity of the cool surge plasma is 
estimated using CM inversions. The opacity decreases from the base of the surge 
to the tip. The decrease in opacity coincides with the increase in the speed of 
the cloud material. The low excitation energy of 10.2~eV of the H$\alpha$ 
transition implies its strong sensitivity to temperature. When cool gas of 
considerable opacity is propelled to higher atmospheric layers it will dilute 
and assume higher temperatures. Thus, ionization of neutral hydrogen becomes 
important. Surprisingly, the location where the surges fades from the H$\alpha$ 
observations coincides with the interface where the surge brightens in 
He\,\textsc{ii} $\lambda$304~\AA.

The H$\alpha$ observations present only a small fraction of the overall data 
covering the two homologous surges, that is, a detailed analysis of H$\alpha$, 
H$\beta$, and Cr\,\textsc{i} spectra along with more advanced CM inversions 
\citep[see][]{Tsiropoula2012} and magnetic field information \citep[see 
e.g.,][for Stokes Inversions based on Response functions (SIR) of 
Stokes-$I$]{delToro1996a} will be presented in a forthcoming study. Such an 
investigation will focus on line formation and peculiar features in spectral 
lines associated with surging plasma. Multi-line spectroscopy with the VTT 
echelle spectrograph has the advantage of co-spatial and co-temporal 
observations, which is beneficial for numerical modelling. The synthesis of 
spectral lines can thus rely on a broad range of physical parameters describing 
the magnetic field and plasma flows in the photosphere and chromosphere. The 
first results presented in Sect.~\ref{SEC36} illustrate the quality and 
potential of CM inversions, which facilitated investigating the coupling between 
photosphere, chromosphere, transition region, and corona and determining the 
response of the upper atmosphere to low-chromospheric reconnection as expected 
for surges \citep[e.g.,][]{Guglielmino2010}.

Optimizing setup and recording settings of the VTT echelle spectrograph enables 
even higher-resolution spectroscopy of strong chromospheric absorption lines 
such as H$\alpha$ and H$\beta$ along with magnetically sensitive photospheric 
lines. In 2019, cadences below one minute were achieved for scans of $100\arcsec 
\times 180\arcsec$. Routine observations with this setup will provide more 
datasets in the future, which are suitable for investigating energetic, 
eruptive, and explosive events in the Sun's lower atmosphere. Concerning surges, 
higher temporal resolution is essential to track phase velocities of dark and 
bright features along the surge trajectory \citep[e.g.,][]{Yang2014} or to 
investigate secondary effects such as filament/prominence oscillations initiated 
by surges \citep[e.g.,][]{Chen2008}. These data are also ideally suited for 
statistical studies and will become publicly available within the framework of 
the EU Horizon 2020 projects SOLARNET and ESCAPE.


\begin{acknowledgements}
The Vacuum Tower Telescope is operated by the German consortium of the 
Leibniz-Institut f\"ur Sonnenphysik (KIS) in Freiburg, the Leibniz-Institut 
f\"ur Astrophysik Potsdam (AIP), and the Max-Planck-Institut f\"ur 
Sonnensystemforschung (MPS) in G\"ottingen. SDO HMI and AIA data are provided by 
the Joint Science Operations Center -- Science Data Processing. This study was 
supported by grants DE~787/5-1 of the Deutsche Forschungsgemeinschaft (DFG) and 
18-08097J of the Czech Science Foundation. In addition, the support by the 
European Commission's Horizon 2020 Program under grant agreements 824064 (ESCAPE 
-- European Science Cluster of Astronomy \& Particle physics ESFRI research 
infrastructures) and 824135 (SOLARNET -- Integrating High Resolution Solar 
Physics) is highly appreciated. ED is grateful for the generous financial 
support from German Academic Exchange Service (DAAD) in form of a doctoral 
scholarship. Many thanks to the referee for insightful comments and suggestions 
on structure and contents, which significantly improved quality and scope of the 
article.
\end{acknowledgements}


\end{document}